\documentclass[aps,pra,amsmath,amssymb,superscriptaddress]{revtex4}
\usepackage{times}
\usepackage{latexsym}
\usepackage{graphicx}
\usepackage{amsmath,amssymb}
\usepackage{tensor}
\usepackage{verbatim,bbm}
\usepackage{epstopdf}
\usepackage{xcolor}
\usepackage{color}
\usepackage[normalem]{ulem}

\DeclareMathOperator{\Tr}{Tr}

\begin{document}

\title{Continuous variable port-based teleportation}

\author{Jason L. Pereira} \email{jason.pereira@fi.infn.it}
\affiliation{ INFN Sezione di Firenze, via G. Sansone 1, I-50019, Sesto Fiorentino (FI), Italy }
\affiliation{Department of Physics and Astronomy, University of Florence,
via G. Sansone 1, I-50019 Sesto Fiorentino (FI), Italy}
\affiliation{Department of Computer Science, University of York, York YO10 5GH, UK}

\author{Leonardo Banchi}
\affiliation{Department of Physics and Astronomy, University of Florence,
via G. Sansone 1, I-50019 Sesto Fiorentino (FI), Italy}
\affiliation{ INFN Sezione di Firenze, via G. Sansone 1, I-50019, Sesto Fiorentino (FI), Italy }

\author{Stefano Pirandola}
\affiliation{Department of Computer Science, University of York, York YO10 5GH, UK}

\date{\today}

\begin{abstract}
    Port-based teleportation is a generalisation of the standard teleportation protocol which does not require unitary operations by the receiver. This comes at the price of requiring $N>1$ entangled pairs, while $N=1$ for the standard teleportation protocol. The lack of correction unitaries allows port-based teleportation to be used as a fundamental theoretical tool to simulate arbitrary channels with a general resource, with applications to study fundamental limits of quantum communication, cryptography and sensing, and to define general programmable quantum computers. Here we introduce a general formulation of port-based teleportation in continuous variable systems and study in detail the $N=2$ case. In particular, we interpret the resulting channel as an energy truncation and analyse the kinds of channels that can be naturally simulated after this restriction.
\end{abstract}

\maketitle

\section{Introduction}

Quantum teleportation is a fundamental protocol in quantum information~\cite{bennett_teleporting_1993,braunstein_teleportation_1998,pirandola_advances_2015}. In its original formulation, it involves perfectly transmitting a qudit using a pre-shared maximally entangled discrete variable (DV) state. A projective Bell measurement is carried out on the state that is to be transmitted (the signal state) and one half of the entangled (resource) state. The measurement result is sent to the receiver, who then carries out one of a set of teleportation unitaries (which one depends on the measurement result) on their half of the resource state to recover the signal state. The concept has since been generalised in a number of ways.

It has been extended to continuous variable (CV) states. First, this was done via the naive approach of simply replacing the maximally entangled DV states with maximally entangled CV states and the measurements with projective measurements onto such states~\cite{vaidman_teleportation_1994}. Such a protocol is not experimentally realisable, however, since a maximally entangled CV state has infinite energy. Instead, the Braunstein-Kimble protocol was developed, which uses finite-energy two-mode squeezed vacuum (TMSV) states~\cite{braunstein_teleportation_1998}. The measurement is of the displacement of a superposition of the signal state and half of the entangled state and the teleportation unitaries, in this case, are displacements. The trade-off is that, for any finite energy resource, the channel enacted is no longer the identity, although it approaches the identity in the asymptotic limit (see Refs.~\cite{pirandola_theory_2018,pirandola_teleportation_2018} for related concepts of strong and uniform convergence in CV teleportation).

Resource states other than the maximally entangled state have been considered. Changing the resource state (without changing the measurement performed) can change the channel enacted (i.e. the transformation applied to the signal state to get the received state). In the standard qudit teleportation case, the channel is simply the identity, but by changing the resource, one can enact any Pauli channel~\cite{bowen_teleportation_2001}. Altering the classical communication stage allows an even wider range of channels to be enacted~\cite{cope_simulation_2017}. This generalisation leads to another use for the teleportation protocol.

Teleportation can be used to send quantum states from one physical location to another. The fact that channels other than the identity can be enacted opens up another possibility. By carrying out teleportation with a resource other than the maximally entangled state, we can apply a channel to a state (generalising the idea of quantum gate teleportation~\cite{gottesman_demonstrating_1999}). This method allows certain channels to be applied to states deterministically (as long as the resource state can be prepared without errors). It can also be used as a particular protocol for channel simulation~\cite{pirandola_fundamental_2017}, a mathematical tool for simulating a quantum channel via local operations and classical communications (LOCC) and a given resource state. Channel simulation allows 
finding computable upper bounds on quantum communication capacities and key rates for a general channel in terms of the entanglement of the resource state~\cite{pirandola_fundamental_2017}, as quantified by 
the relative entropy of entanglement~\cite{vedral_quantifying_1997,vedral_entanglement_1998}.
Note that this is a theoretical technique for calculating information
theoretic quantities of the channel that relies only on the possibility of
enacting the channel via teleportation, rather than a practical application of the teleportation protocol.
One limitation of using teleportation to implement channels or for channel simulation is that only Pauli channels can be enacted via standard qudit teleportation with a modified resource. Indeed, the measurement-dependent post-processing by the receiver forces a requirement, teleportation covariance \cite{pirandola_theory_2018,cope_simulation_2017}, which in turn implies that the only channels that can be enacted by a teleportation protocol are those that commute with the post-processing teleportation unitaries.

Ishizaka and Hiroshima introduced a new type of teleportation protocol, called port-based teleportation~\cite{ishizaka_asymptotic_2008,ishizaka_quantum_2009}, later studied and improved in a number of works~\cite{studzinski_port-based_2017,mozrzymas_optimal_2018,kabgyun_generalization_2020,christandl_asymptotic_2021,pereira_characterising_2021,Studzinski_square-root_2022}. In port-based teleportation (PBT), the sender and receiver share several entangled states (called ports), rather than one, and these collectively form the shared resource state. The measurement on the sender's half of the resource state and the signal state gives one of $N$ possible results, where $N$ is the number of ports (for the deterministic version of the protocol). Crucially, the only post-processing required is to select the correct port, based on the measurement result. The lack of teleportation unitaries on the receiver state allows for the simulation of any channel, without the requirement of teleportation covariance. However, unlike in the standard teleportation case, for any finite number of ports this protocol does not enact perfect simulation, namely an identity channel on the signal state, even with the optimal resource state and the optimal measurement. However, the enacted channel approaches the identity asymptotically in the number of ports. The lack of post-processing is a key advantage, as it means every qudit channel can be enacted to arbitrary precision with enough ports and the correct resource state. 
Aside from finding upper bounds on quantum capacities, other applications include finding 
theoretical limits for quantum sensing \cite{pirandola_fundamental_2019} and defining 
programmable quantum computers \cite{banchi_convex_2020}, where the resource state defines the program.
One important basic case uses a tensor product of $N$ maximally entangled qudits as the resource and the square root measurement, also called pretty good measurements \cite{hausladen_pretty_1994,hausladen_classical_1996}.

The Braunstein-Kimble protocol has the same limitation as standard DV teleportation of only being able to enact channels that commute with the teleportation unitaries, namely the set of displacement operators~\cite{pirandola_theory_2018}. One particular example of an entire class of channels that cannot be simulated at all are channels that apply an energy truncation. Specifically, consider the set of all channels that accept any input but only output states with an average photon number below some number $E$. For any such channel, we can always choose an input with a very high energy so that the outputs of the ``simulated" channel and the teleportation channel are arbitrarily far apart. This limitation exists for any resource state. Note, however, that it is possible to modify the Braunstein-Kimble protocol slightly so that it is able to simulate channels that apply an energy truncation by, for instance, restricting the set of post-processing displacements by imposing a maximum displacement magnitude.

In this paper, we will consider a natural follow-on to the idea of PBT: namely, whether the concept can be extended to CV systems. We call this CV-PBT. Here, we avoid the naive approach of replacing maximally entangled qudit ports with maximally entangled but unphysical CV ports. Instead, we use finite energy TMSV ports as the resource, and construct our square root measurement in a similar way (from finite energy TMSVs). 

Continuous variable PBT is largely absent from the literature, except from Boiselle's master thesis~\cite{boisselle_jason_port-based_2014}, 
so it is worth addressing how this work differs from ours. Boisselle proposes (in Chapter~6) the following protocol for teleporting coherent states starting from the same type of resource as us (a number of finite energy TMSVs). The parties first carry out entanglement purification on their shared resource to replace the (CV) TMSV states with a smaller number of maximally entangled DV states. The sender then encodes the (CV) coherent state in a multi-qubit state (thereby applying an energy truncation) and then sends each qubit via standard DV-PBT (using the purified resource). Finally, the receiver reconstructs (an truncated approximation of) the original coherent state.
Whilst this is technically using PBT to teleport an initially CV state, this is quite far from what one might expect from a CV extension of PBT. Fundamentally, it is DV-PBT with some pre- and post-processing to convert both the resource and the signal to DV states. The final state is DV, not CV, since the reconstruction process uses beamsplitters to recombine the multi-qubit state into a single mode, but cannot undo the truncation. In contrast, our protocol uses a CV resource with a CV measurement, and the output states can have support over the entire (overcomplete) basis of coherent states.

We will propose the general form of the CV-PBT protocol, but will only calculate explicit expressions for the channels enacted in the two and three port cases, with particular focus on the two port case. Whilst this is a somewhat limited result, it serves two main purposes. Firstly, it is a proof in principle of a CV extension to PBT, showing that we can get meaningful results using this new protocol, and that it does (imperfectly) transmit a quantum state. Secondly, we find that the protocol enacts an energy truncation on the transmitted state, so that any input state results in an output state with bounded energy. This means that CV-PBT can simulate a completely, qualitatively different set of channels than standard CV teleportation, even in the two port case. Finally, we have developed the $N$ mode formalism, detailing the process of calculating the channel output for more than three modes, but without giving explicit expressions.

\section{Specification of the protocol}

Let us begin by setting out the protocol in its general, $N$ port form, before focusing more narrowly on the two port case. Suppose a sender, Alice, is trying to transmit a one-mode CV state to a receiver, Bob, using a pre-shared resource state, and without sending any quantum states from one party to the other. Denote Alice's half of the shared resource collectively as system $A$, with the individual ports constituting systems $A_1$, $A_2$, etc. Bob's half of the shared resource constitutes system $B$ (with individual ports constituting systems $B_i$). The signal state is system $C$.

The initial resource state, $\phi^{[N]}_x$, consists of $N$ TMSVs, each with a squeezing specified by $x$. We write
\begin{equation}
    \phi^{[N]}_x = \bigotimes_{i=1}^{N} S(\lambda_x) \left|00\middle>\middle<00\right|_{A_i B_i} S(\lambda_x)^{\dagger},
\end{equation}
where $x$ is a squeezing parameter ($\lambda_x=\tanh x$) and $S(\lambda)$ is the two-mode squeezing operator, which acts on the zero state as
\begin{equation}
    S(\lambda) \left|00\right> = \sqrt{1-\lambda^2} \sum_{n=0}^{\infty}(-\lambda)^n \left|nn\right>.
\end{equation}

Next, we must specify the measurement. We define
\begin{equation}
    \sigma_{i,y}^{[N]} = S(\lambda_y) \left|00\middle>\middle<00\right|_{A_i C} S(\lambda_y)^{\dagger} \otimes \mathcal{I}_{A_{\Bar{i}}B},
\end{equation}
where $i$ is any element of the set of integers from $1$ to $N$, $\Bar{i}$ is the same set excluding $i$, and $\mathcal{I}$ denotes the identity. Then define $\rho_{y}^{[N]} = \sum_{i=1}^N \sigma_{i,y}^{[N]}$. Henceforth, we neglect the superscript $[N]$ where not required. We now construct a POVM with elements
\begin{equation}
    M^{(i)}_{y} = \rho_{y}^{-\frac{1}{2}}\sigma_{i,y}\rho_{y}^{-\frac{1}{2}} + \frac{1}{N} \{\rho_{y}\}_K,
\end{equation}
where $\{X\}_K$ denotes the kernel of $X$ (i.e. the subspace spanned by those eigenvectors of $X$ that have eigenvalues of zero). By construction, the elements sum to the identity, so this is a valid measurement. Note, however, that even for DV systems, this type of (square root) measurement is, in practice, experimentally difficult.

Alice carries out this measurement on systems $A$ and $C$ (the signal state and her half of the shared resource), then sends the result, $j$ (an integer between $1$ and $N$), to Bob. Bob then picks and retains port $j$ (i.e. system $B_j$), whilst discarding the rest of his subsystems.

Denoting the channel enacted by CV-PBT on an initial state $\psi$ as $\mathcal{E}_{x,y}$, we can write $\mathcal{E}_{x,y}[\psi] = N \Tr_{AB_{\bar{i}C}}[M^{(i)}_{y} \psi\otimes\phi^{[N]}_x]$, where - due to the permutation symmetry of the resource and the measurement operators - it does not matter what value $i$ takes. To characterise the effect of the channel, we could choose to express the output for an (arbitrary) coherent state input. This serves as a complete characterisation of a CV channel because any CV state can be expressed as a pseudo-probability distribution over the set of coherent states (and due to the linearity of quantum channels).

\subsection{CV-PBT channel for two ports}

Before going through the details, we will give a high level overview of how we go about finding an explicit expression for the channel output for a coherent state input, $\mathcal{E}_{x,y}[\left|\alpha_{\mathrm{coh}}\middle>\middle<\alpha_{\mathrm{coh}}\right|]$ (where we denote the output of the teleportation channel, for an input state $\psi$, as $\mathcal{E}_{x,y}[\psi]$.). We start by writing an eigenvector decomposition of $\rho_y$. We then calculate $\rho_{y}^{-\frac{1}{2}}\sigma_{i,y}\rho_{y}^{-\frac{1}{2}}$ (i.e. the part of the measurement that lies in the support of $\rho_y$), and hence $M^{(i)}$. We apply this expression to calculate $\mathcal{E}_{x,y}[\left|a\middle>\middle<b\right|]$, the effect of the channel on an arbitrary entry in the number state basis. Finally, we use the number state expression for a coherent state to calculate $\mathcal{E}_{x,y}[\left|\alpha_{\mathrm{coh}}\middle>\middle<\alpha_{\mathrm{coh}}\right|]$.

For $N=2$, we can decompose $\rho$ in terms of its eigenvectors as
\begin{equation}
    \rho_y = \sum_{m=0}^{\infty} \left[ \Xi_{y,m}^{+}\left|\Psi_{y,m}^+\middle>\middle<\Psi_{y,m}^+\right| + \Xi_{y,m}^{-}\left|\Psi_{y,m}^-\middle>\middle<\Psi_{y,m}^-\right| \right],\label{eq: rho decomp}
\end{equation}
where the eigenvalues $\{\Xi_{y,m}^{\pm}\}$ and the eigenstates $\{\left|\Psi_{y,m}^{\pm}\right>\}$ are defined as
\begin{equation}
    \Xi_{y,m}^{\pm} = 1\pm\chi_{y,m},\quad
    \left|\Psi_{y,m}^{\pm}\right> = \sqrt{\frac{1-\lambda_y^2}{2\Xi_{y,m}^{\pm}}} \sum_{n=0}^{\infty} (-\lambda_y)^n \left|n\right>_C \otimes (\left|nm\right> \pm \left|mn\right>)_{A},\label{eq: eigenstates}
\end{equation}
and where $\chi_{x,r}$ is given by $\chi_{x,r} = (1-\lambda_x^2)\lambda_x^{2r}$. The details of this eigendecomposition are given in Appendix~\ref{app: eigendecomposition}.

Defining $\sigma'_{i,y} = \rho_{y}^{-\frac{1}{2}}\sigma_{i,y}\rho_{y}^{-\frac{1}{2}}$, we get (see Appendix~\ref{app: measurements})
\begin{equation}
    \sigma'_{1,y} = \frac{1}{2} \sum_{m=0}^{\infty} (\left|\Psi_{y,m}^{+}\right> + \left|\Psi_{y,m}^{-}\right>)(\left<\Psi_{y,m}^{+}\right| + \left<\Psi_{y,m}^{-}\right|),\quad
    \sigma'_{2,y} = \frac{1}{2} \sum_{m=0}^{\infty} (\left|\Psi_{y,m}^{+}\right> - \left|\Psi_{y,m}^{-}\right>)(\left<\Psi_{y,m}^{+}\right| - \left<\Psi_{y,m}^{-}\right|).\label{eq: signal decomp}
\end{equation}
Including the contribution from the kernel of $\rho$, we get
\begin{equation}
    M_y^{(1)}=\frac{\mathcal{I}}{2} + \frac{1}{2}\sum_{m=0}^{\infty} \left(\left|\Psi_{y,m}^{+}\middle>\middle<\Psi_{y,m}^{-}\right| + \left|\Psi_{y,m}^{-}\middle>\middle<\Psi_{y,m}^{+}\right|\right),\quad
    M_y^{(2)}=\frac{\mathcal{I}}{2} - \frac{1}{2}\sum_{m=0}^{\infty} \left(\left|\Psi_{y,m}^{+}\middle>\middle<\Psi_{y,m}^{-}\right| + \left|\Psi_{y,m}^{-}\middle>\middle<\Psi_{y,m}^{+}\right|\right).\label{eq: measurement decomp}
\end{equation}

Now let us calculate the effect of the protocol on an arbitrary entry in the number state basis. Recalling that we denote the resource state with squeezing parameter $x$ by $\phi_x$, we define $\phi^{(i)}_{ab,x} = \Tr_{B_{\Bar{i}}}[\left|a\middle>\middle<b\right|_C \otimes\phi_x]$. If we have measurement outcome $i$, we are interested in $\Tr_{AC}[M^{(i)}_{y} \phi^{(i)}_{ab,x}]$. Without loss of generality, we set $i=1$. We express $\phi^{(1)}_{ab,x}$ as
\begin{equation}
    \phi^{(1)}_{ab,x} = (1-\lambda_x^2) \left|a\middle>\middle<b\right|_C \otimes \left( \sum_{p,q=0}^{\infty}(-\lambda_x)^{p+q} \left|pp\middle>\middle<qq\right|_{A_1 B_1}\right) \otimes \left( \sum_{r=0}^{\infty} \chi_{x,r} \left|r\middle>\middle<r\right|_{A_2} \right).\label{eq: arbitrary number state}
\end{equation}
Since we have traced over mode $B_2$, mode $A_2$ is in a thermal state. We calculate (see Appendix~\ref{app: coherent state})
\begin{equation}
    \mathcal{E}_{x,y}[\left|a\middle>\middle<b\right|] =
    (1-\lambda_x^2)(1-\lambda_y^2)\lambda_y^{a+b} \sum_{m=0}^{\infty} \frac{\chi_{x,a+m}}{\sqrt{1-\chi_{y,m}^2}}(\lambda_x^{b-a}\left|a\middle>\middle<b\right| - \delta_{ab}\left|m\middle>\middle<m\right|)
    + \delta_{ab}\sum_{m=0}^{\infty} \chi_{x,m} \left|m\middle>\middle<m\right|.\label{eq: channel on ab}
\end{equation}
Note that if $a=b$, the trace of Eq.~(\ref{eq: channel on ab}) is $1$, verifying that we have a valid channel.

A generic CV state can be expressed as a combination of coherent states via the so-called P-representation 
\cite{vogel2006quantum,sudarshan_equivalence_1963,glauber_coherent_1963}
\begin{equation}
	\rho = \int d\alpha \,P(\alpha)\, |{\alpha_{\mathrm{coh}}}\rangle\!\langle\alpha_{\mathrm{coh}}|,
	\label{eq: p repr}
\end{equation}
where $P(\alpha)$ is a quasi-probability distribution, namely $P(\alpha)$ may be negative for 
non-classical states. The possibility of expressing a general CV state as a combination of 
diagonal elements as in Eq.~\eqref{eq: p repr} comes from the fact that coherent states 
form an overcomplete basis. Note however that Eq.~\eqref{eq: p repr} does not 
represent a convex combination of coherent states, as $P(\alpha)$ may be negative. 
Thanks to the P-representation, we can characterise any channel via its action on 
coherent states. 

Mathematically, a generic coherent state can be expressed as
\begin{equation}
    \left|\alpha_{\mathrm{coh}}\middle>\middle<\alpha_{\mathrm{coh}}\right| = \exp[-|\alpha|^2] \sum_{a,b=0}^{\infty} \frac{\alpha^{a}\alpha^{*b}}{\sqrt{a!b!}}\left|a\middle>\middle<b\right|,
\end{equation}
where $\alpha^*$ is the complex conjugate of $\alpha$. Defining $\Omega_{x,y} = \sum_{m=0}^{\infty} \frac{\chi_{x,m}}{\sqrt{1-\chi_{y,m}^2}}$, we can characterise the channel by writing
\begin{equation}
    \begin{split}
        \mathcal{E}_{x,y}[\left|\alpha_{\mathrm{coh}}\middle>\middle<\alpha_{\mathrm{coh}}\right|] =&
        e^{-\left(1 - \lambda_x^2 \lambda_y^2\right)|\alpha|^2}(1-\lambda_x^2)(1-\lambda_y^2)\Omega_{x,y} \left|\lambda_x\lambda_y\alpha_{\mathrm{coh}}\middle>\middle<\lambda_x\lambda_y\alpha_{\mathrm{coh}}\right|\\
        &+ \sum_{m=0}^{\infty} \chi_{x,m}\left(1-e^{-\left(1 - \lambda_x^2 \lambda_y^2\right)|\alpha|^2}\frac{(1-\lambda_x^2)(1-\lambda_y^2)}{\sqrt{1-\chi_{y,m}^2}}\right) \left|m\middle>\middle<m\right|.
    \end{split}\label{eq: 2 port characterisation}
\end{equation}
Thanks to the P-representation \eqref{eq: p repr}, this is a complete characterisation of the channel. Note from Eq.~(\ref{eq: channel on ab}) that there is no global phase applied to the transmitted state, so if we have an idler state, teleportation will not lead to a relative phase between the idler and the teleported state. Note too that if $|\alpha|$ is close to $0$ and $\lambda_y$ is sufficiently small (specifically, if $(1-\lambda_y^2)^{-2}-1<(1-\lambda_x^2)^{2}$), the $m=0$ term in the final sum in Eq.~(\ref{eq: 2 port characterisation}) can be negative and the first term can be $>1$. This is not unphysical, since coherent states and number states are not orthonormal, so the contributions cancel out, but it can complicate calculations, so it is often easier to work in what we call the positive regime (as opposed to the negative regime), for which $(1-\lambda_y^2)^{-2}-1\geq(1-\lambda_x^2)^{2}$.

\section{Properties of the teleportation channel}

\subsection{Effect of the channel}

By looking at the form of Eq.~(\ref{eq: 2 port characterisation}), we can gain a more physically intuitive understanding of the effect of the channel. One observation we can immediately make is that the channel is phase-insensitive. With some probability, the channel output is the same as that of a lossy channel with a transmission of $\lambda_x^2\lambda_y^2$ (which we denote by $\mathcal{L}_{x,y}$). Otherwise, the output is a thermal state (with average photon number $\frac{\lambda_x^2}{1-\lambda_x^2}$) with some non-Gaussian corrections, which are diagonal in the number state basis. It should be noted, however, that even without these non-Gaussian corrections, the channel output would be a convex combination of two Gaussian states, which is not itself Gaussian. We can rewrite Eq.~(\ref{eq: 2 port characterisation}) as
\begin{equation}
    \begin{split}
        \mathcal{E}_{x,y}[\left|\alpha_{\mathrm{coh}}\middle>\middle<\alpha_{\mathrm{coh}}\right|] =
        e^{-(1-\tau_{x,y})|\alpha|^2}g_{x,y}\Omega_{x,y} \left|\sqrt{\tau_{x,y}}\alpha_{\mathrm{coh}}\middle>\middle<\sqrt{\tau_{x,y}}\alpha_{\mathrm{coh}}\right| + (1-e^{-(1-\tau_{x,y})|\alpha|^2}g_{x,y})\rho^{\mathrm{th}}_{x}\\
        + e^{-(1-\tau_{x,y})|\alpha|^2}g_{x,y}\sum_{m=0}^{\infty} \chi_{x,m}\left(1-(1-\chi_{y,m}^2)^{-\frac{1}{2}}\right) \left|m\middle>\middle<m\right|,
    \end{split}\label{eq: 2 port characterisation rearranged}
\end{equation}
where $\rho^{\mathrm{th}}_{x}$ is the thermal state with coefficients $\chi_{x,m}$, $\tau_{x,y}=\lambda_x^2 \lambda_y^2$, and $g_{x,y}=(1-\lambda_x^2)(1-\lambda_y^2)$. The magnitude of the non-Gaussian corrections (the last term in Eq.~(\ref{eq: 2 port characterisation rearranged})) is smaller for higher number states. In the case of $m=0$, $\chi_{x,m}\left(1-(1-\chi_{y,m}^2)^{-\frac{1}{2}}\right)=(1-\lambda_x^2)\left(1-\lambda_y^{-1}(2-\lambda_y^{2})^{-\frac{1}{2}}\right)$, whilst as $m\to\infty$, $\chi_{x,m}\left(1-(1-\chi_{y,m}^2)^{-\frac{1}{2}}\right)\to 0$.

\begin{figure}[tb]
\vspace{+0.1cm}
\centering
\includegraphics[width=1\linewidth]{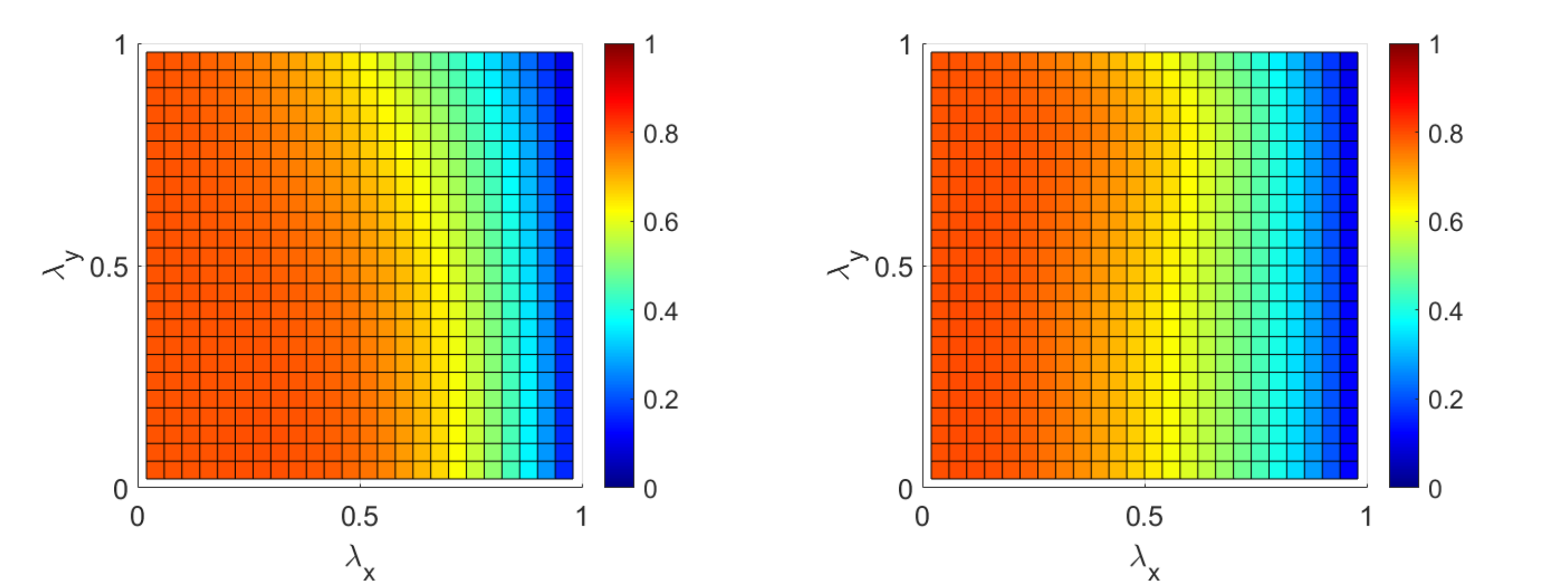}\caption{The input-output fidelity when one half of a TMSV state is acted on with CV-PBT. The plot on the left shows the two port case, whilst the plot on the right shows the three port case.}
\label{fig: tmsv}
\end{figure}

In Fig.~(\ref{fig: tmsv}), we show the input-output fidelity when one half of a TMSV with squeezing parameter $\lambda_{\mathrm{in}}=\frac{1}{3}$ is teleported via CV-PBT. This squeezing parameter corresponds to the signal state having an average photon number of $\frac{1}{2}$. We also show the input-output fidelity for three mode CV-PBT. The calculations were carried out using numerical methods, as detailed in Appendix~\ref{app: TMSV}, using MATLAB code, which is available as supplementary material.

\subsection{Energy of the channel output}

Let us consider the energy of the channel output for a coherent state input. The average photon number of a coherent state is $|\alpha|^2$, whilst for our output, it is
\begin{equation}
    E_{|\alpha|^2}=\sum_{n=0}^{\infty} n \left<n\right|\mathcal{E}_{x,y}[\left|\alpha_{\mathrm{coh}}\middle>\middle<\alpha_{\mathrm{coh}}\right|]\left|n\right> =
    e^{-\left(1 - \tau_{x,y}\right)|\alpha|^2}g_{x,y}\left(\tau_{x,y}\Omega_{x,y}|\alpha|^2 
    - \sum_{m=0}^{\infty} m\frac{\chi_{x,m}}{\sqrt{1-\chi_{y,m}^2}}\right) + \frac{\lambda_x^2}{1-\lambda_x^2}.\label{eq: energy output coherent}
\end{equation}
A key realisation here is that, for any finite $x$, this quantity is bounded, even as $|\alpha|\to\infty$. In fact, the maximum output energy for any input state is (see Appendix~\ref{app: output energy})
\begin{equation}
    E_{\mathrm{max}} = \frac{\tau_{x,y}g_{x,y}\Omega_{x,y}}{1 - \tau_{x,y}}\exp\left[-\left(1+\frac{1 - \tau_{x,y}}{\tau_{x,y}\Omega_{x,y}}\sum_{m=0}^{\infty} m\frac{\chi_{x,m}}{\sqrt{1-\chi_{y,m}^2}}\right)\right]+\frac{\lambda_x^2}{1-\lambda_x^2}.\label{eq: energy output max}
\end{equation}

The channel therefore applies an energy truncation. It is also of interest that this energy truncation is not a ``hard cut-off" (i.e. a truncation in the number state basis, which would map CV states to DV), but rather a constraint on the average photon number of the output. The outputs of CV-PBT remain CV states and their support continues to be the entire set of coherent states. Unlike for a hard cut-off, there is no quantum channel that applies only this kind of truncation whilst leaving the state otherwise undisturbed (i.e. there is no channel that acts as the identity on all states with an average photon number less than some maximum value but not on states with a higher energy). This follows from the linearity of quantum channels and of the energy of a state, by considering a convex combination of a low energy state and a high energy state. This means that $\mathcal{E}_{x,y}$ cannot be decomposed into the pointwise application of a channel that does not apply an energy truncation and a channel that just applies an energy truncation.

\begin{figure}[tb]
\vspace{+0.1cm}
\centering
\includegraphics[width=0.5\linewidth]{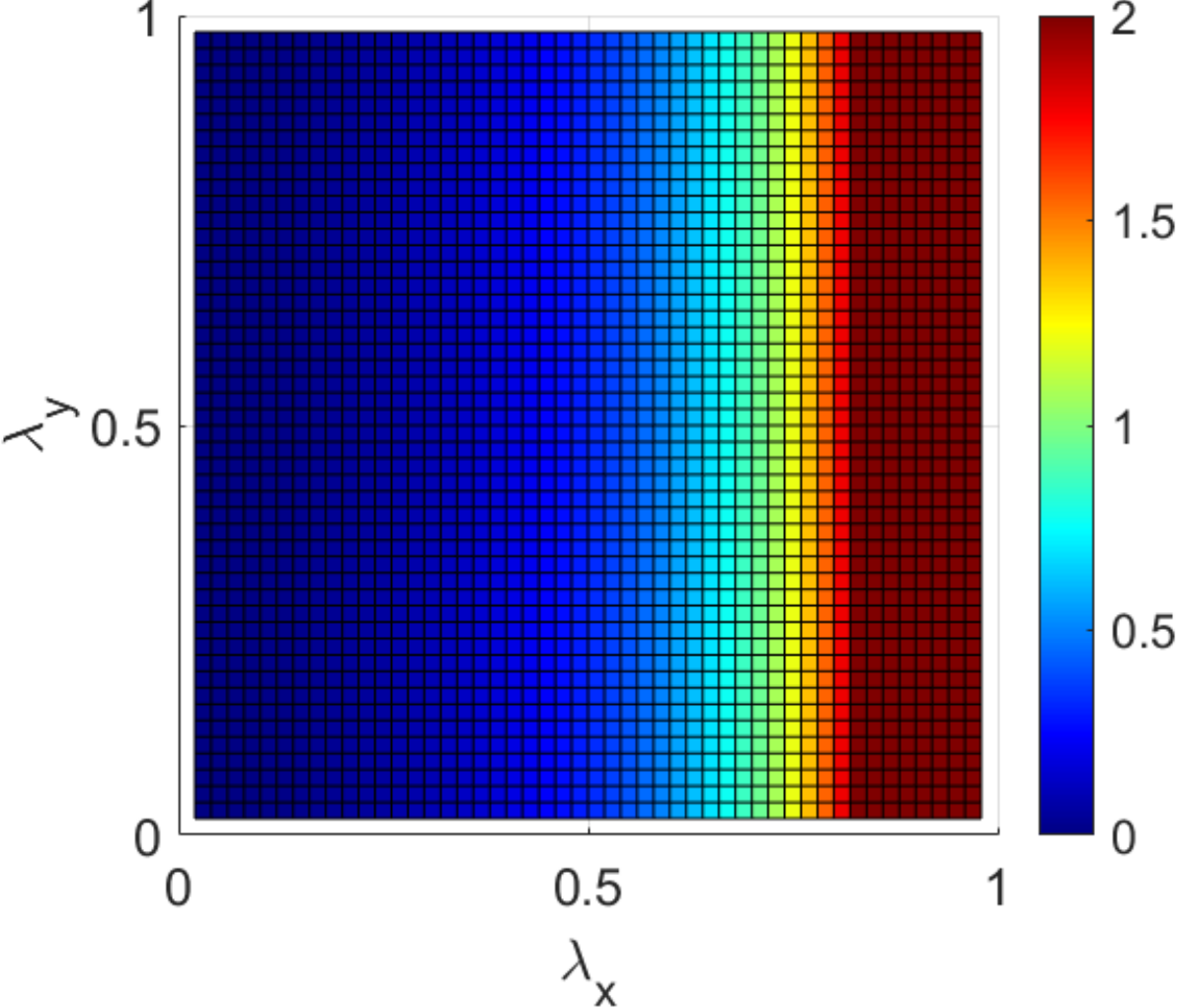}\caption{The maximum energy (average photon number) of the output state for two port CV-PBT. Note that the colour bar ends at $2$, however the energy continues to increase beyond this value.}
\label{fig: energy}
\end{figure}

We plot the maximum output energy against the parameters $\lambda_x$ and $\lambda_y$ in Fig.~\ref{fig: energy}. As can be seen, the maximum output energy grows with $\lambda_x$, whilst having almost no dependence on $\lambda_y$. This is expected, since $\lambda_x$ determines the energy of resource state, whilst $\lambda_y$ only controls the particular measurement carried out upon it by the sending party.

\subsection{Distance from physically relevant channels}

It is unlikely that a physical scenario can be modelled exactly by the channel given in Eq.~(\ref{eq: 2 port characterisation rearranged}). If we want to use CV-PBT for channel simulation, we must ask: what physically relevant channels does Eq.~(\ref{eq: 2 port characterisation rearranged}) resemble? When we say that two channels are similar to each other, we mean that if the same signal state is the input for both channels, the resulting output states will be close to each other (generally in the sense of the trace norm between them). One important metric for assessing the similarity of two channels is the diamond norm. The diamond norm is the trace norm between the output states maximised over all possible input states (including those with idler modes).

Since CV-PBT applies a lossy channel with a probability that depends on the energy of the input, one simple channel we might consider comparing it to is a lossy channel with the same loss, which we will call $\mathcal{L}_{x,y}$. Since lossy channels do not apply an energy truncation, for sufficiently energetic inputs, the output of $\mathcal{L}_{x,y}$ can be arbitrarily far away from the output of $\mathcal{E}_{x,y}$, so the diamond norm between the two channels will be $2$ (the maximum value). Instead, we can consider the energy constrained diamond norm (first introduced in Ref.~\cite{pirandola_fundamental_2017} for the study of the two-way assisted capacities of bosonic channels and then generalised in Refs.~\cite{shirokov_energy-constrained_2018,winter_energy-constrained_2017}). Instead of maximising over every possible input state, we only maximise over those states for which the energy of the signal state is less than or equal to some maximum value. We define
\begin{equation}
    D_{x,y,E}=\sup_{\psi_{IS}\in\Psi_E} \|\mathcal{I}\otimes\mathcal{E}_{x,y}[\psi] - \mathcal{I}\otimes\mathcal{L}_{x,y}[\psi]\|,\quad
    \Psi_E = \{\psi_{IS}|\Tr[\hat{n}_S\psi]\leq E\},
\end{equation}
where $S$ denotes the mode that is sent through the channel (the signal mode), $I$ denotes an idler system, and $\hat{n}_S$ is the photon number operator on the signal mode.

\begin{figure}[tb]
\vspace{+0.1cm}
\centering
\includegraphics[width=0.5\linewidth]{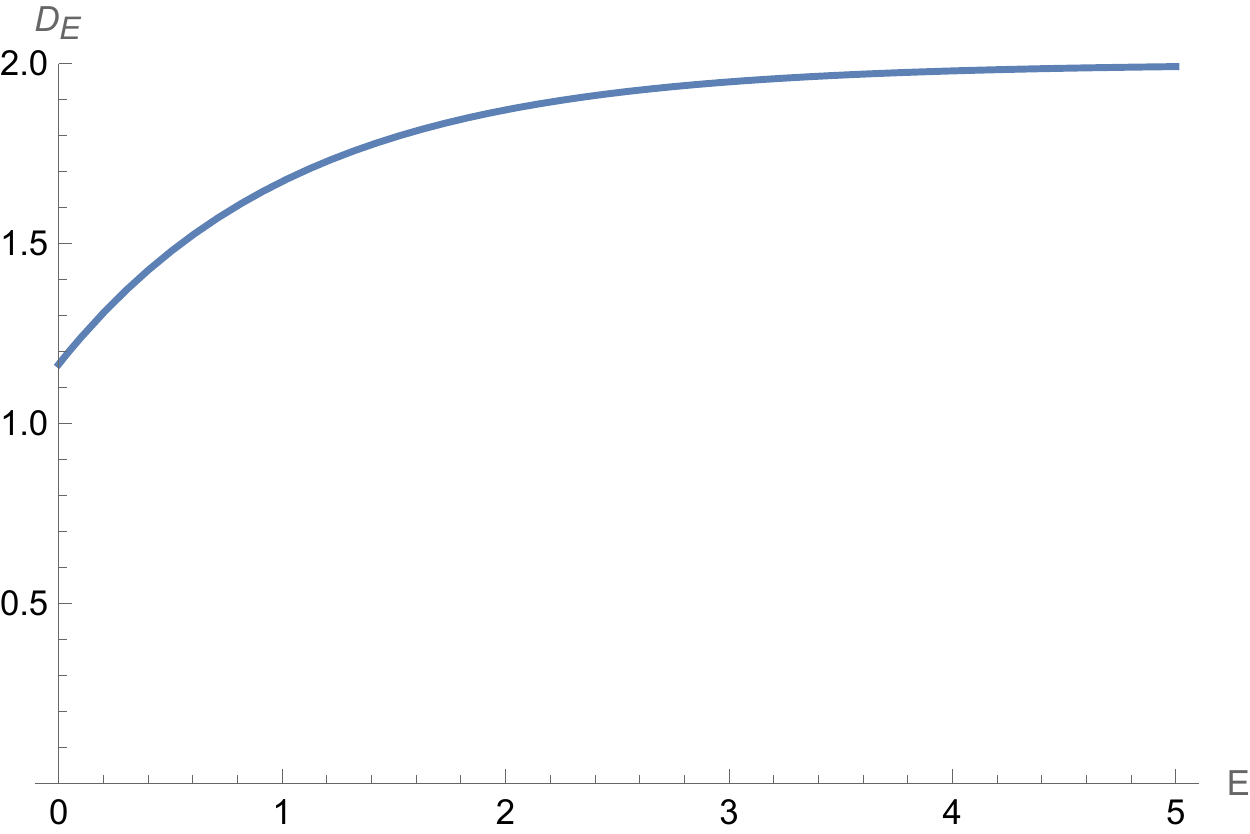}\caption{Upper bound on the energy constrained diamond norm between the two port CV-PBT channel with $\lambda_x=0.5$ and $\lambda_y=0.5$ and a lossy channel with a transmissivity of $\tau=0.0625$. $E$ is the average number of photons for the input state.}
\label{fig: lossy}
\end{figure}

We find (see Appendix~\ref{app: lossy channel comp}) that, for the positive regime,
\begin{equation}
    D_{x,y,E}[\mathcal{E}_{x,y},\mathcal{L}_{x,y}] \leq 2\left(1 - e^{-E \left(1 - \tau_{x,y}\right)}g_{x,y}\Omega_{x,y}\right).\label{eq: lossy channel comp bound}
\end{equation}
This bound takes a more complicated form in the negative regime. It is not necessarily a good bound, since it relies on the triangle inequality. It is illustrated, for $\lambda_x=0.5$ and $\lambda_y=0.5$, in Fig.~\ref{fig: lossy}. The plot shows that the lossy channel is not very similar to the two port CV-PBT channel for these parameter values, even at low energies, since even for $\alpha=0$ the bound on the energy constrained diamond norm is $\sim 1.16$ and the true value (since in this case it is simple to calculate exactly, by setting $\alpha=0$ in Eq.~(\ref{eq: 2 port characterisation})) is $\sim 0.94$.

We might instead consider channels of the form
\begin{equation}
    \mathcal{C}[\left|\alpha_{\mathrm{coh}}\middle>\middle<\alpha_{\mathrm{coh}}\right|]=e^{-\kappa |\alpha|^2} f \left|\sqrt{\tau}\alpha_{\mathrm{coh}}\middle>\middle<\sqrt{\tau}\alpha_{\mathrm{coh}}\right| + (1-e^{-\kappa |\alpha|^2} f) \rho^{\mathrm{th}}_{h},\label{eq: channel model better}
\end{equation}
where $\kappa$, $f$, $\tau$, and $h$ are parameters to be specified. We will call this type of channel an energy-dependent replacement channel, since it enacts a replacement channel with a probability that depends on the energy of the input state. At first glance, this is a less simple and generally useful type of channel than the lossy channel, however we can conceive of physical scenarios for which it could be a good model. We will present one such physical situation, purely as an example.

Suppose we have a filament of material that we wish to send laser pulses through (either to probe it or to transmit quantum information). Any photonic state passing through the material has a chance to be consumed in an interaction that replaces it with a thermal state (e.g. they could be absorbed by reactive sites distributed through the material that then randomly emit a photon from a thermal distribution), whilst otherwise it emerges subject to some damping (e.g. due to reflection at the boundaries). Suppose its chance of being consumed in such an interaction depends on the time spent within the filament of material, so that its probability of not interacting is $e^{\frac{nL}{c}}$, where $n$ is the refractive index of the filament, $L$ is its length, and $c$ is the speed of light in a vacuum. The probability of interacting is then $1-e^{\frac{nL}{c}}$. Finally, suppose that the filament has a large second-order non-linear refractive index, so that the refractive index, $n$, is dependent on the intensity, $I$, of the optical state, and can be written as $n_0+n_2 I$. $I$ is proportional to the average photon number of the state. With these ingredients in place, we can see that we can model this situation using the quantum channel given in Eq.~(\ref{eq: channel model better}).

We emphasise that, whilst this is a very specific physical scenario, we are not interested in any particular physical scenario, but rather offer this as one example of a quantum channel that could be simulated by our teleportation protocol. We could also calculate the diamond norm between the teleportation channel and other channel models, or could change the teleportation channel significantly by changing the resource state.

Assuming the positive regime and setting $\kappa=1-\tau_{x,y}$, $f=g_{x,y}\Omega_{x,y}$, $\tau=\tau_{x,y}$, and $h=x$ in Eq.~(\ref{eq: channel model better}), we compare the resulting energy-dependent replacement channel, $\mathcal{C}_{x,y}$, to $\mathcal{E}_{x,y}$. The diamond norm is (see Appendix~\ref{app: edrc comp})
\begin{equation}
    D_{x,y}[\mathcal{E}_{x,y},\mathcal{C}_{x,y}] = 2g_{x,y}\sum_{m=0}^{m_c} \chi_{x,m}\left((1-\chi_{y,m_c}^2)^{-\frac{1}{2}}-\Omega_{x,y}\right),\label{eq: edrc diamond}
\end{equation}
where we define $m_c$ as the largest integer for which $(1-\chi_{y,m_c}^2)^{-\frac{1}{2}}>\Omega_{x,y}$. This can be much smaller than the bound on the energy constrained diamond norm between the same channel ($\mathcal{E}_{x,y}$) and $\mathcal{L}_{x,y}$, even for small energy constraints, although this is not surprising since Eq.~(\ref{eq: edrc diamond}) is exact and since we have specifically chosen the form of Eq.~(\ref{eq: channel model better}) to be similar to the PBT channel.

\section{Channel simulation example}

Suppose we have a physical scenario that can be modelled by one of two channels of the form
\begin{equation}
    \mathcal{C}[\left|\alpha_{\mathrm{coh}}\middle>\middle<\alpha_{\mathrm{coh}}\right|]=e^{-(1-\tau_{x,y}) |\alpha|^2} g_{x,y}\Omega_{x,y} \left|\sqrt{\tau_{x,y}}\alpha_{\mathrm{coh}}\middle>\middle<\sqrt{\tau_{x,y}}\alpha_{\mathrm{coh}}\right| + (1-e^{-(1-\tau_{x,y}) |\alpha|^2} g_{x,y}\Omega_{x,y}) \rho^{\mathrm{th}}_{x},\label{eq: example channel}
\end{equation}
where $\tau_{x,y}$, $g_{x,y}$, and $\Omega_{x,y}$ are the previously defined functions of $\lambda_x$ and $\lambda_y$, $\lambda_y=2^{-\frac{1}{4}}$, and $\lambda_{x^{\pm}}=2^{-\frac{1}{4}}\pm \frac{1}{2}\delta$. We want to send probe states through the channel and then carry out a final measurement, in order to determine which of the channels we have. The scenario could involve the filaments described in the previous section or could be some other situation with a similar mathematical description.

Suppose we want to upper bound the distinguishability (diamond norm) of this pair of channels after a single channel use. At first glance, this could be a tricky problem, since even for a coherent state input, the difference between the outputs is a complicated combination of two different coherent states and two different thermal states. It is not obvious how one would diagonalise this difference to calculate the trace norm of the difference, let alone maximise over all possible P-distributions. If we allowed multiple channel uses, this problem would become even more difficult, since we must account for all possible processing operations between uses. For a finite dimensional system, we could solve the problem numerically via semidefinite programming, but since we are looking at CV states, this is only possible if we apply some truncation.

\begin{figure}[tb]
\vspace{+0.1cm}
\centering
\includegraphics[width=0.5\linewidth]{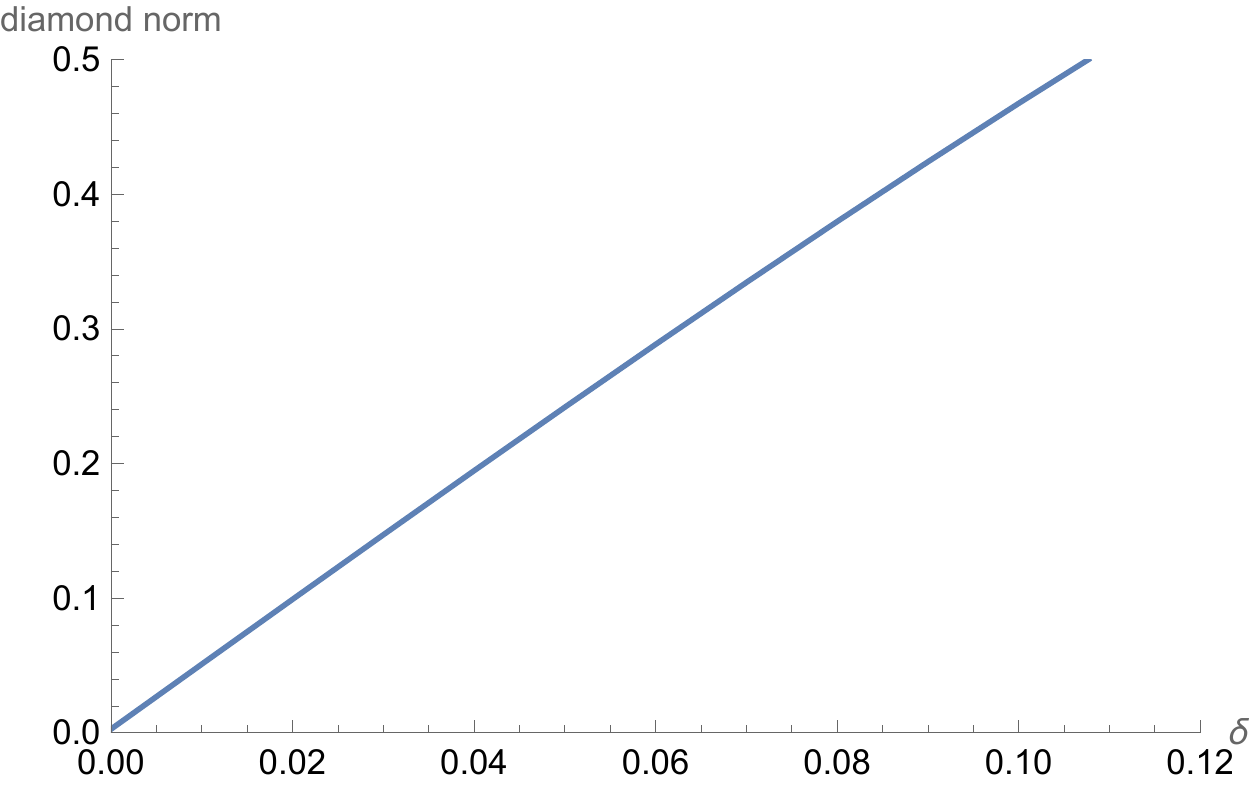}\caption{Upper bound on the diamond norm between two channels of the form given in Eq.~(\ref{eq: example channel}), derived using simulation with CV-PBT.}
\label{fig: sim}
\end{figure}

Alternatively, we can simulate the channels using CV-PBT with the same values of $\lambda_x$ and $\lambda_y$ (these are not necessarily the optimal values that give the tightest bounds, but we choose them for convenience). For more details about how channel simulation can be used to bound the distinguishability of two channels, see Ref.~\cite{pirandola_fundamental_2019}. Using Eq.~(\ref{eq: edrc diamond}) and bounding the trace norm between the resource states $\phi_{x_{\pm}}$ using the fidelity and the Fuchs-van de Graaf inequality, we get the plot in Fig.~\ref{fig: sim}, which bounds the diamond norm between the channels in terms of $\delta$. The extension to multiple channel uses is simple. Note that it is not possible to simulate channels of this type using the Braunstein-Kimble protocol.

\section{CV-PBT for $N$ ports}

For the $N$ port case, we do not give an explicit expression for the channel output, but rather show how one can go about calculating the output state. Recall that the action of the PBT channel is $\mathcal{E}_{x,y}[\psi] = N \Tr_{AB_{\bar{1}C}}[M^{(1)}_{y} \psi\otimes\phi^{[N]}_x]$. We do not present, as for the two port case, an explicit expression for the measurement operator in terms of $\lambda_y$, but instead give a method for finding it. This requires quite a lot of new notation, which we introduce very quickly here, but which is explained in more detail in Appendix~\ref{app: N ports}.

The eigenstates of $\rho_y$ take the form
\begin{equation}
    \left|\psi(\mathcal{M},i)\right>_{CA} = \sqrt{\frac{1-\lambda_y^2}{\Xi^{(i)}_{\mathcal{M}}}}\sum_{n=0}^{\infty} (-\lambda_y)^n\left|n\right>_C \otimes \sum_{\Phi\in\mathcal{P}} \eta^{(i)}_{\mathcal{M},\Phi} \left(\Phi\left|n\right>\left|\mathcal{M}\right>\right)_A,\label{eq: eigenstate gen}
\end{equation}
where $\Xi^{(i)}_{\mathcal{M}}$ is the corresponding eigenvalue, given by
\begin{equation}
    \Xi^{(i)}_{\mathcal{M}} = 1-(1-\lambda_y^2)\sum_{m\in \bar{\mathcal{M}}}\lambda_y^{2m} + h(\mathcal{M},\eta^{(i)}_{\mathcal{M}}),\quad
    h(\mathcal{M},\eta^{(i)}_{\mathcal{M}}) = \frac{1-\lambda_y^2}{\eta^{(i)}_{\mathcal{M},\Phi}}\sum_{m\in \bar{\mathcal{M}}} \lambda_y^{2m} \sum_{\Phi'\in\mathcal{P}_m} \eta^{(i)}_{\mathcal{M},\Phi\cdot\Phi'}.
\end{equation}
$\mathcal{M}$ is a multiset consisting of $N-1$ integers $\geq 0$ (i.e. a set but with repetition allowed), $\Phi$ is a permutation of the $A$ modes (i.e. a way of rearranging $\left|n\right>\left|\mathcal{M}\right>$), $\mathcal{P}$ is the set of all such (non-degenerate) permutations, $\mathcal{P}_m$ is the set of all permutations of the $A$ modes that only exchange $n$ and a mode with value $m$, and $\tilde{\mathcal{P}}$ is the subset of $\mathcal{P}$ that only acts on the last $N-1$ modes. $\eta^{(i)}_{\mathcal{M}}$ is a set of parameter values that defines a particular eigenvector (the $i$-th eigenvector corresponding to the multiset $\mathcal{M}$), and finding the eigendecomposition of $\rho_y$ is equivalent to finding all allowed sets of parameters $\eta_{\mathcal{M}}$. This is a very brief introduction to the formalism we have developed for the $N$ port case; for further details, see Appendix~\ref{app: N ports}.

The measurement operator can now be expressed as
\begin{equation}
    M_{y}^{(1)} = \frac{1}{N}\mathcal{I} + \sum_{\mathcal{M}} \sum_{\eta_{\mathcal{M}}^{(\alpha)},\eta_{\mathcal{M}}^{(\beta)}} (1-\delta_{\alpha\beta}) \Big( \sum_{\Phi\in\Tilde{\mathcal{P}}}\eta_{\mathcal{M},\Phi}^{(\alpha)*}\eta_{\mathcal{M},\Phi}^{(\beta)} \Big) \left|\psi(\mathcal{M},\alpha)\middle>\middle<\psi(\mathcal{M},\beta)\right|,\label{eq: measurement N modes}
\end{equation}
where the first sum is over all distinct multisets of $N-1$ integers ($\geq 0$) and the second sum is over all eigenvectors forming a basis for $\rho_y$. We define the following function of $\mathcal{M}$:
\begin{equation}
    \Gamma(\mathcal{M}) = \sum_{\eta_{\mathcal{M}}^{(\alpha)},\eta_{\mathcal{M}}^{(\beta)}} \frac{1-\delta_{\alpha\beta}}{\sqrt{\Xi_{\mathcal{M}}^{(\alpha)}\Xi_{\mathcal{M}}^{(\beta)}}} \Big( \sum_{\Phi\in\Tilde{\mathcal{P}}}\eta_{\mathcal{M},\Phi}^{(\alpha)*}\eta_{\mathcal{M},\Phi}^{(\beta)} \Big) \eta_{\mathcal{M}}^{(\beta)\dagger}\eta_{\mathcal{M}}^{(\alpha)},\quad
    \Gamma(\mathcal{M},\Phi',\Phi'') = \Gamma(\mathcal{M})_{\Phi',\Phi''},\label{eq: gamma expr}
\end{equation}
where $\Gamma(\mathcal{M})$ is a Hermitian matrix and $\Gamma(\mathcal{M},\Phi',\Phi'')$ refers to a specific element of this matrix. So long as we can find all of the parameter sets $\eta^{(i)}_{\mathcal{M}}$ for a specific multiset $\mathcal{M}$, it is simple to calculate $\Gamma(\mathcal{M})$ for that multiset. Finally, in Eqs.~(\ref{eq: a neq b gen expr N modes}) and (\ref{eq: a=b gen expr N modes}), we give simple expressions to calculate the channel output for a given input state from sums of specific elements of $\Gamma(\mathcal{M})$ summed over all multisets $\mathcal{M}$.

The only remaining difficulty in calculating the channel output for $N$ port PBT is therefore in finding the set $\{\eta^{(i)}_{\mathcal{M}}\}$ for every $N-1$ element multiset $\mathcal{M}$. In Appendix~\ref{app: N port eigendecomposition}, we show how the eigenvectors of $\rho_y$ can be found. This is simple for $N=3$, but quickly becomes difficult to do analytically for large $N$. 

In Appendix~\ref{app: 3 ports}, we apply the $N$ port formalism and derive explicit expressions for the channel output for an arbitrary input element of the number state basis. These expressions are found in Eqs.~(\ref{eq: n port output breakdown}), (\ref{eq: mm a neq b output}), (\ref{eq: mm a=b output}), (\ref{eq: lm a neq b output}), and (\ref{eq: lm a=b output}). The equations consist of sums of elements of $\Gamma(\mathcal{M})$, as defined in Eq.~(\ref{eq: gamma expr}), however Eqs.~(\ref{eq: lm vectors}), (\ref{eq: lm eigenvalues}), (\ref{eq: gamma mm}), and (\ref{eq: gamma lm}) allow it to be constructed analytically for every multiset $\mathcal{M}$.

Finally, in Appendix~\ref{app: N port numerics}, we explain how one could numerically approximate the channel output, by only summing over a finite number of multisets $\mathcal{M}$. This approach is also illustrated, for the two and three port cases, in Appendix~\ref{app: TMSV} and in the supplementary MATLAB code.

\section{Discussion}

We have generalised the PBT protocol by introducing a CV version that can be carried out using finite energy resources (and hence is physically achievable). Unlike other CV teleportation protocols, the only post-processing required is a swap operation between modes.

We have explicitly calculated the resulting teleportation channel for the two and three port cases, demonstrating in principle that it is feasible to analytically calculate the teleportation channel for CV-PBT. In the $N$ port case, we have developed a method by which the channel output can, in principle, be calculated numerically. The teleportation channel in the two port case has a maximum output energy, and so imposes an energy constraint, but without a hard cut-off in the number state basis. This opens up a new class of channels that can be simulated using teleportation.

A possible extension of this work is simplifying the formalism for the $N$ port case in such a way that it is possible to give an explicit analytical expression for the channel output in the general case. Another possibility is to investigate how changing the resource state changes the teleportation channel. In this work we have generalised deterministic PBT protocols to the CV case, however there exists another type of PBT protocol (in the DV case). In probabilistic PBT, the protocol has a chance of failing, but if it succeeds, the teleportation is perfect. Further research could generalise the probabilistic PBT protocol in a similar way to our generalisation of the deterministic version. Finally, in the DV case ``optimal" PBT optimises over both the measurement and the resource state to maximise the closeness of the teleportation channel to an identity channel, whilst retaining the property that no post-processing is required. Future research could consider the effect of optimising the measurement to decrease the energy constrained diamond norm between the teleportation channel and a lossy channel in a similar way.

\smallskip
\begin{acknowledgments}
J.~L.~P. and S.~P acknowledge funding from the European Union's Horizon 2020 Research and Innovation Action under grant agreement No. 862644 (FET-OPEN project: Quantum readout techniques and technologies, QUARTET). J.~L.~P. and L.~B. acknowledge funding from the U.S. Department of Energy, Office of Science, National Quantum Information Science Research Centers, Superconducting Quantum Materials and Systems Center (SQMS) under the contract No. DE-AC02-07CH11359.
\end{acknowledgments}

\appendix

\section{Eigendecomposition of $\rho_y$}\label{app: eigendecomposition}

We calculate the eigendecomposition by considering the effect of $\rho_y$ on the state $\left|pqr\right>_{C A_1 A_2}$.
\begin{equation}
    \rho_y \left|pqr\right> = (1-\lambda_y^2)(-\lambda_y)^p \sum_{s=0}^{\infty}(-\lambda_y)^s (\delta_{pq} \left|ssr\right> + \delta_{pr} \left|sqs\right>)\label{eq: rho gen effect 2 ports}
\end{equation}
where $\delta$ is the Kronecker delta. It is clear that any state of the form $\left|pqr\right>$ for which $p$ is not equal to at least one of $q$ and $r$ lies in the kernel of $\rho_s$.

We construct a generic state
\begin{equation}
    \left|\psi(\eta,\gamma)\right> = \sum_{m,n=0}^{\infty} \left|n\right>_C \otimes (\eta_{mn}\left|nm\right> + \gamma_{mn}\left|mn\right>)_{A_1 A_2},\label{eq: gen state 2 ports}
\end{equation}
where the functions $\eta$ and $\gamma$ define a specific state. Any state with no component lying in the kernel of $\rho_y$ must be of this form. $\rho_y$ acts on this state as
\begin{equation}
    \begin{split}
        \rho_y\left|\psi(\eta,\gamma)\right> = (1-\lambda_y^2) \sum_{q,m=0}^{\infty} (-\lambda_y)^{q}\left|q\right>\otimes \left[\left(\sum_{n=0}^{\infty}(-\lambda_y)^{n}\eta_{mn}+(-\lambda_y)^{m}\gamma_{mm}\right)\left|qm\right>\right.\\
        \left.+ \left(\sum_{n=0}^{\infty}(-\lambda_y)^{n}\gamma_{mn}+(-\lambda_y)^{m}\eta_{mm}\right)\left|mq\right>\right],
    \end{split}
\end{equation}
where we note the extra contribution due to the state $\left|mmm\right>$.

We can now construct the following (necessary and sufficient) conditions for the state to be an eigenvector:
\begin{align}
    &\Xi(1-\lambda_y^2)^{-1}(-\lambda_y)^{-q}\eta_{mq} = \sum_{n=0}^{\infty}(-\lambda_y)^{n}\eta_{mn}+(-\lambda_y)^{m}\gamma_{mm}\quad\forall m, q\neq m,\label{eq: cond eta}\\
    &\Xi(1-\lambda_y^2)^{-1}(-\lambda_y)^{-q}\gamma_{mq} = \sum_{n=0}^{\infty}(-\lambda_y)^{n}\gamma_{mn}+(-\lambda_y)^{m}\eta_{mm}\quad\forall m, q\neq m,\label{eq: cond gamma}\\
    &\Xi(1-\lambda_y^2)^{-1}(-\lambda_y)^{-m}(\eta_{mm}+\gamma_{mm}) = \sum_{n=0}^{\infty}(-\lambda_y)^{n}(\eta_{mn}+\gamma_{mn})+(-\lambda_y)^{m}(\eta_{mm}+\gamma_{mm})\quad\forall m,\label{eq: cond mmm}
\end{align}
where $\Xi$ is an eigenvalue. Note that the right-hand sides of Eqs.~(\ref{eq: cond eta}) and (\ref{eq: cond gamma}) have no $q$-dependence. We can therefore write
\begin{equation}
    \eta_{mq} = (-\lambda_y)^{q} c_{\eta}^{(m)},\quad\gamma_{mq} = (-\lambda_y)^{q} c_{\gamma}^{(m)}\quad\forall m, q\neq m,
\end{equation}
where $c_{\eta}^{(m)}$ and $c_{\gamma}^{(m)}$ are constant for a fixed value of $m$. Substituting these expressions back into Eqs.~(\ref{eq: cond eta}) and (\ref{eq: cond gamma}) and carrying out the sum over $n$ (excluding the $n=m$ term), we get
\begin{align}
    &\Xi=g_m + h_{m,c},\\
    &g_m = 1-(1-\lambda_y^2)\lambda^{2m},\quad
    h_{m,c} = \frac{1-\lambda^2}{c_{\eta}^{(m)}}(-\lambda_y)^{m}(\eta_{mm}+\gamma_{mm}) =
    \frac{1-\lambda^2}{c_{\gamma}^{(m)}}(-\lambda_y)^{m}(\eta_{mm}+\gamma_{mm}).
\end{align}
We can rewrite Eq.~(\ref{eq: cond mmm}) as $\Xi(-\lambda_y)^{-m}(\eta_{mm}+\gamma_{mm}) = (g_m + h_{m,c})(c_{\eta}^{(m)}+c_{\gamma}^{(m)})$, and dividing both sides by $\Xi$, we get
\begin{equation}
    (-\lambda_y)^{-m}(\eta_{mm}+\gamma_{mm}) = c_{\eta}^{(m)}+c_{\gamma}^{(m)}.
\end{equation}
We can therefore choose, without loss of generality, to set
$\eta_{mm} = (-\lambda_y)^{m} c_{\eta}^{(m)}$, $\gamma_{mm} = (-\lambda_y)^{m} c_{\gamma}^{(m)}$.

All eigenvectors can therefore be written in the form
\begin{equation}
    \left|\psi(m,c_{\eta},c_{\gamma})\right> = \sum_{n=0}^{\infty} (-\lambda_y)^{n}\left|n\right>_C \otimes (c_{\eta}\left|nm\right> + c_{\gamma}\left|mn\right>)_{A_1 A_2},
\end{equation}
where we no longer sum over $m$, because any two vectors of this form but with different values of $m$ are orthogonal. It only remains to determine for which values of $c_{\eta}$ and $c_{\gamma}$ we get valid eigenvectors. Our expression for $h_{m,c}$ becomes
\begin{equation}
    h_{m,c} = \frac{1-\lambda^2}{c_{\eta}}\lambda_y^{2m}(c_{\eta}+c_{\gamma}) =
    \frac{1-\lambda^2}{c_{\gamma}}\lambda_y^{2m}(c_{\eta}+c_{\gamma}),
\end{equation}
and clearly this can only be satisfied if $c_{\eta}=c_{\gamma}$ or if $c_{\eta}=-c_{\gamma}$. Finally, by imposing the normalisation condition, the eigenvalues and eigenvectors given in Eq.~(\ref{eq: eigenstates}) follow directly.

\section{Calculation of POVM elements}\label{app: measurements}

Consider the effect of $\sigma_{1/2,y}$ on the state $\left|pqr\right>_{C A_1 A_2}$:
\begin{align}
    &\sigma_{1,y}\left|pqr\right>_{C A_1 A_2} = \delta_{pq}(1-\lambda_y^2)\sum_{s=0}^{\infty} (-\lambda_y)^{p+s}\left|ssr\right>_{C A_1 A_2},\label{eq: 2 port sigma on gen}\\
    &\sigma_{2,y}\left|pqr\right>_{C A_1 A_2} = \delta_{pr}(1-\lambda_y^2)\sum_{s=0}^{\infty} (-\lambda_y)^{p+s}\left|sqs\right>_{C A_1 A_2}.
\end{align}
Using these expressions, we can calculate
\begin{equation}
    \sigma_{1,y}\left|\Psi_{y,m}^{\pm}\right> = \frac{(1-\lambda_y^2)^{\frac{3}{2}}}{\sqrt{2\Xi_{y,m}^{\pm}}} \left( \sum_{r=0}^{\infty} \lambda_y^{2r} \pm \lambda_y^{2m} \right) \sum_{n=0}^{\infty} (-\lambda_y)^n \left|nnm\right>
    = \sqrt{\frac{(1-\lambda_y^2)\Xi_{y,m}^{\pm}}{2}} \sum_{n=0}^{\infty} (-\lambda_y)^n \left|nnm\right>,
\end{equation}
with a corresponding expression for $\sigma_{2,y}$. Next, we find
\begin{align}
    &\left<\Psi_{y,m}^{\pm}\right|\sigma_{1,y}\left|\Psi_{y,m}^{\pm}\right> = \frac{(1-\lambda_y^2)}{2} \left( \sum_{n=0}^{\infty} (-\lambda_y)^n (\left<nnm\right| \pm \left<nmn\right|) \right)\sum_{p=0}^{\infty} (-\lambda_y)^p \left|ppm\right>
    = \frac{\Xi_{y,m}^{\pm}}{2},\\
    &\left<\Psi_{y,m}^{\mp}\right|\sigma_{1,y}\left|\Psi_{y,m}^{\pm}\right> = \frac{(1-\lambda_y^2)}{2}\sqrt{\frac{\Xi_{y,m}^{\pm}}{\Xi_{y,m}^{\mp}}} \left( \sum_{n=0}^{\infty} (-\lambda_y)^n (\left<nnm\right| \mp \left<nmn\right|) \right)\sum_{p=0}^{\infty} (-\lambda_y)^p \left|ppm\right>
    = \frac{\sqrt{\Xi_{y,m}^{+}\Xi_{y,m}^{-}}}{2}.
\end{align}
Combining these expressions (and the corresponding expressions for $\sigma_{2,y}$) with Eq.~(\ref{eq: rho decomp}), we get Eq.~(\ref{eq: signal decomp}).

We note that the kernel of $\rho_y$ can be expressed as
\begin{equation}
    \{\rho_{y}\}_K = \mathcal{I} - \sum_{m=0}^{\infty} \left[ \left|\Psi_{y,m}^+\middle>\middle<\Psi_{y,m}^+\right| + \left|\Psi_{y,m}^-\middle>\middle<\Psi_{y,m}^-\right| \right],
\end{equation}
and so Eq.~(\ref{eq: measurement decomp}) follows immediately. Explicitly, $M_y^{(1)}$ takes the form
\begin{equation}
    M_y^{(1)}=\frac{\mathcal{I}}{2} + \frac{1-\lambda_y^2}{2}\sum_{p,q=0}^{\infty} (-\lambda_y)^{p+q} \left|p\middle>\middle<q\right| \otimes \sum_{m=0}^{\infty} \frac{\left|pm\middle>\middle<qm\right|-\left|mp\middle>\middle<mq\right|}{\sqrt{\Xi_{y,m}^{+}\Xi_{y,m}^{-}}},\label{eq: M1 explicit}
\end{equation}
with a similar expression for $M_y^{(2)}$. Finally, note that $\Xi_{y,m}^{+}\Xi_{y,m}^{-}=1-\chi_{y,m}^2$.

\section{Calculating the effect of the channel on a coherent state}\label{app: coherent state}

Since the resource state is identical for both ports, both measurement outcomes result in the same output state. We can therefore calculate $\Tr_{AC}[M_y^{(1)} \phi^{(i)}_{ab,x}]$ and then double the result. Using Eq.~(\ref{eq: M1 explicit}), we get
\begin{equation}
    2\Tr_{AC}[M_y^{(1)} \phi^{(i)}_{ab,x}] = \delta_{ab}\sum_{m=0}^{\infty}\chi_{x,m}\left|m\middle>\middle<m\right|+(1-\lambda_x^2)(1-\lambda_y^2)(-\lambda_y)^{a+b}\sum_{m=0}^{\infty}\frac{(-\lambda_x)^{a+b}\chi_{x,m}\left|a\middle>\middle<b\right| -\delta_{ab} \lambda_x^{2m}\chi_{x,a}\left|m\middle>\middle<m\right|}{\sqrt{1-\chi_{y,m}^2}}.
\end{equation}
From the definitions of $\chi$ and $\lambda$, we can write $\chi_{x,a}\lambda_x^{2b}=\chi_{x,a+b}$, and so can simplify the previous expression, getting Eq.~(\ref{eq: channel on ab}).

It is worth verifying that $\mathcal{E}_{x,y}[\left|a\middle>\middle<a\right|]$ is a valid quantum state. From Eq.~(\ref{eq: channel on ab}), we have
\begin{equation}
    \mathcal{E}_{x,y}[\left|a\middle>\middle<a\right|] =
    (1-\lambda_x^2)(1-\lambda_y^2)\lambda_y^{2a} \sum_{m=0}^{\infty} \frac{\chi_{x,a+m}}{\sqrt{1-\chi_{y,m}^2}}(\left|a\middle>\middle<a\right| - \left|m\middle>\middle<m\right|)
    + \sum_{m=0}^{\infty} \chi_{x,m} \left|m\middle>\middle<m\right|.\label{eq: channel on aa}
\end{equation}
It is clear that the trace of the first term is $0$ (and so the trace of the entire state is $1$), but it is not immediately obvious that the state is a positive operator. Since our expression is in diagonal form (in the number state basis), proving positivity amounts to showing that the following inequality holds for all $m$ and for any $a$ (except for $m=a$):
\begin{equation}
    \chi_{x,m}(1-(1-\lambda_x^2)(1-\lambda_y^2)\lambda_x^{2a}\lambda_y^{2a}(1-\chi_{y,m}^2)^{-\frac{1}{2}})\geq 0.
\end{equation}
After some rearrangement, this becomes
\begin{equation}
    \frac{1-\chi_{y,m}^2}{(1-\lambda_y^2)^2}=(1-\lambda_y^2)^{-2} - \lambda_y^{4m} \geq (1-\lambda_x^2)^2\lambda_x^{4a}\lambda_y^{4a}.
\end{equation}
For $m\geq 1$, the left-hand side is $\geq 1$, whilst the right-hand side is always $\leq 1$. On the other hand, if $m=0$, the left-hand side becomes $\cosh^4 y - 1$ whilst the right-hand side is $\leq \lambda_y^{4a}$, and so the inequality holds for any $a\geq 1$. If $m=a$, the first term in Eq.~(\ref{eq: channel on aa}) cancels out. Hence, Eq.~(\ref{eq: channel on aa}) describes a valid quantum state, as expected.

Applying Eq.~(\ref{eq: channel on ab}), we see that CV-PBT applied to a coherent state gives the output
\begin{equation}
    \begin{split}
        \mathcal{E}_{x,y}[\left|\alpha_{\mathrm{coh}}\middle>\middle<\alpha_{\mathrm{coh}}\right|] =
        \exp[-|\alpha|^2](1-\lambda_x^2)(1-\lambda_y^2) \Bigg[ \sum_{a,b=0}^{\infty} \frac{(\lambda_x \lambda_y)^{a+b}\alpha^a\alpha^{*b}}{\sqrt{a!b!}}\left(\sum_{m=0}^{\infty} \frac{\chi_{x,m}}{\sqrt{1-\chi_{y,m}^2}}\right)\left|a\middle>\middle<b\right|\\
        - \sum_{m=0}^{\infty} \frac{\lambda_x^{2m}}{\sqrt{1-\chi_{y,m}^2}}\left(\sum_{a=0}^{\infty} \chi_{x,a}\frac{|\lambda_y\alpha|^{2a}}{a!}\right)\left|m\middle>\middle<m\right| \Bigg]
        + \sum_{m=0}^{\infty} \chi_{x,m} \left|m\middle>\middle<m\right|.
    \end{split}
\end{equation}
By explicit evaluation,
\begin{equation}
    \sum_{a=0}^{\infty} \chi_{x,a}\frac{|\lambda_y\alpha|^{2a}}{a!} = \exp[|\alpha|^2 \lambda_x^2 \lambda_y^2] \cosh^{-2} x,
\end{equation}
and consequently,
\begin{equation}
    \begin{split}
        \mathcal{E}_{x,y}[\left|\alpha_{\mathrm{coh}}\middle>\middle<\alpha_{\mathrm{coh}}\right|] =
        \exp[-|\alpha|^2](1-\lambda_x^2)(1-\lambda_y^2) \Bigg[ \sum_{a,b=0}^{\infty} \frac{(\lambda_x \lambda_y)^{a+b}\alpha^a\alpha^{*b}}{\sqrt{a!b!}}\left(\sum_{m=0}^{\infty} \frac{\chi_{x,m}}{\sqrt{1-\chi_{y,m}^2}}\right)\left|a\middle>\middle<b\right|\\
        - \exp[|\alpha|^2 \lambda_x^2 \lambda_y^2]\sum_{m=0}^{\infty} \frac{\chi_{x,m}}{\sqrt{1-\chi_{y,m}^2}}\left|m\middle>\middle<m\right| \Bigg]
        + \sum_{m=0}^{\infty} \chi_{x,m} \left|m\middle>\middle<m\right|.
    \end{split}
\end{equation}
This simplifies to Eq.~(\ref{eq: 2 port characterisation}).

\section{Maximum output energy}\label{app: output energy}

From Eq.~(\ref{eq: energy output coherent}), we know $E_{|\alpha|^2}$, the output energy for an input coherent state with an average photon number of $|\alpha|^2$. To maximise over all inputs, we use the fact that all CV states can be represented by P-distributions. Any one mode state can be written as
\begin{equation}
    \psi = \int P(\alpha) \left|\alpha_{\mathrm{coh}}\middle>\middle<\alpha_{\mathrm{coh}}\right| d^2 \alpha,\label{eq: 1 mode p dist}
\end{equation}
where $P(\alpha)$ is a real function that can take both positive and negative values and where we use $d^2$ to indicate that we are integrating over both the real and imaginary parts of a complex number. Although $P(\alpha)$ can take negative values,
\begin{equation}
    \int P(\alpha) d^2 \alpha = 1.
\end{equation}

Defining $\alpha = |\alpha| e^{i\phi}$, we can express $E(\psi)$, the output energy for an input state $\psi$, as
\begin{equation}
    E(\psi) = \int \int p(|\alpha|)P(\phi| |\alpha|) E_{|\alpha|^2} d\phi~d|\alpha|,\label{eq: energy P dist}
\end{equation}
where we have decomposed $P(\alpha)$ into a product of $p(|\alpha|)$ and the conditional P-distribution $P(\phi| |\alpha|)$. Whilst the conditional P-distribution can again be negative and integrates to $1$ for all values of $|\alpha|$, $p(|\alpha|)$ is a true probability distribution. This is guaranteed by the fact that $|\alpha|^2$ is the average photon number, which is an observable.

Since $E_{|\alpha|^2}$ is not a function of $\phi$, Eq.~(\ref{eq: energy P dist}) becomes
\begin{equation}
    E(\psi) = \int_{0}^{\infty} p(|\alpha|) E_{|\alpha|^2} d|\alpha|.
\end{equation}
We can show that $E_{|\alpha|^2}$ has a (single) maximum, so the maximum possible output energy is achieved if $p(|\alpha|)$ is a delta function centred on this maximum value.

The expression for $E_{|\alpha|^2}$ is of the form
\begin{equation}
    E_{x} = e^{-x\beta}(\gamma x - \delta) + \epsilon,\label{eq: form of energy eq}
\end{equation}
where all of the variables are positive (semidefinite). Differentiating this expression, we get
\begin{equation}
    \frac{d E_{x}}{dx} = e^{-x\beta}(\gamma+\beta\delta-\beta\gamma x),
\end{equation}
which is equal to $0$ for only one $x$-value: $x_{\mathrm{max}} = \frac{\gamma+\beta\delta}{\beta\gamma}$. Using this value, we get
\begin{equation}
    E_{x_{\mathrm{max}}} = \frac{\gamma}{\beta}e^{-(1+\frac{\beta\delta}{\gamma})} + \epsilon.
\end{equation}
Finally, returning to our original variables (i.e. by comparing Eqs.~(\ref{eq: energy output coherent}) and (\ref{eq: form of energy eq})), we recover Eq.~(\ref{eq: energy output max}).

\section{Bounding the energy constrained diamond norm from a lossy channel}\label{app: lossy channel comp}

Defining $\Delta_{\alpha}=\mathcal{E}_{x,y}[\left|\alpha_{\mathrm{coh}}\middle>\middle<\alpha_{\mathrm{coh}}\right|] - \left|\sqrt{\tau_{x,y}}\alpha_{\mathrm{coh}}\middle>\middle<\sqrt{\tau_{x,y}}\alpha_{\mathrm{coh}}\right|$, we evaluate
\begin{equation}
    \Delta_{\alpha} =
    \sum_{m=0}^{\infty} \chi_{x,m}\left(1-e^{-\left(1 - \tau_{x,y}\right)|\alpha|^2}\frac{g_{x,y}}{\sqrt{1-\chi_{y,m}^2}}\right)\left|m\middle>\middle<m\right|
    - \left(1 - e^{-\left(1 - \tau_{x,y}\right)|\alpha|^2}g_{x,y}\Omega_{x,y}\right)\left|\sqrt{\tau_{x,y}}\alpha_{\mathrm{coh}}\middle>\middle<\sqrt{\tau_{x,y}}\alpha_{\mathrm{coh}}\right|.\label{eq: delta expr}
\end{equation}

Recalling that any two mode state can be written as
\begin{equation}
    \psi_{IS} = \int_{\alpha} \int_{\beta} P(\alpha,\beta) \left|\beta_{\mathrm{coh}}\middle>\middle<\beta_{\mathrm{coh}}\right|_I\otimes\left|\alpha_{\mathrm{coh}}\middle>\middle<\alpha_{\mathrm{coh}}\right|_S d^2 \alpha~d^2 \beta,\label{eq: 2 mode p dist}
\end{equation}
where $P(\alpha,\beta)$ is a real function that can take both positive and negative values and where we use $d^2$ to indicate that we are integrating over both the real and imaginary parts of a complex number. Although $P(\alpha,\beta)$ can take negative values,
\begin{equation}
    \int_{\alpha} \int_{\beta} P(\alpha,\beta) d^2 \alpha~d^2 \beta = 1.
\end{equation}
If $P(\alpha,\beta)$ were a true probability distribution, we could calculate the trace norm of Eq.~(\ref{eq: delta expr}) and then use the convexity of the trace norm to bound the trace norm for any input state. However, since $P(\alpha,\beta)$ can take negative values, the convexity argument does not hold. Instead, we define
\begin{equation}
    \alpha = r_{\alpha} e^{i\phi_{\alpha}},\quad
    P(\alpha,\beta)=P(r_{\alpha},\phi_{\alpha},\beta)=P(r_{\alpha})P(\phi_{\alpha},\beta|r_{\alpha}),
\end{equation}
where $r_{\alpha}\geq 0$ and we have decomposed $P(\alpha,\beta)$ into a product of $P(r_{\alpha})$ and the conditional P-distribution $P(\phi_{\alpha},\beta|r_{\alpha})$. Whilst the conditional P-distribution can again be negative and integrates to $1$ for all values of $r_{\alpha}$, $P(r_{\alpha})$ is a true probability distribution. This is guaranteed by the fact that $r_{\alpha}$ corresponds to the observable $\hat{n}_S$ (specifically, $\hat{n}_S=\hat{r}_{\alpha}^2$). Consequently, if we can write an upper bound on
\begin{equation}
    T_{r_{\alpha}} = \left\|\int_{\phi_{\alpha}} \int_{\beta} P(\phi_{\alpha},\beta|r_{\alpha}) \Delta_{\alpha,S}\otimes \left|\beta_{\mathrm{coh}}\middle>\middle<\beta_{\mathrm{coh}}\right|_I d\phi_{\alpha}~d^2\beta\right\|
\end{equation}
that holds for all conditional probability distributions $P(\phi_{\alpha},\beta|r_{\alpha})$, we can then use the convexity of the trace norm over $P(r_{\alpha})$ to bound the trace norm for any given input state.

We can split $\Delta_{\alpha}$ into two contributions:
\begin{align}
    &\Delta_{\alpha}^{\mathrm{diag}} = \sum_{m=0}^{\infty} \chi_{x,m}\left(1-e^{-r_{\alpha}^2 \left(1 - \tau_{x,y}\right)}g_{x,y}(1-\chi_{y,m}^2)^{-\frac{1}{2}}\right)\left|m\middle>\middle<m\right|,\label{eq: diag cont}\\
    &\Delta_{\alpha}^{\mathrm{coh}} = -\left(1 - e^{-r_{\alpha}^2 \left(1 - \tau_{x,y}\right)}g_{x,y}\Omega_{x,y}\right)\left|\sqrt{\tau_{x,y}}\alpha_{\mathrm{coh}}\middle>\middle<\sqrt{\tau_{x,y}}\alpha_{\mathrm{coh}}\right|\label{eq: coherent cont},
\end{align}
where $\Delta_{\alpha}^{\mathrm{coh}}$ is the part of $\Delta_{\alpha}$ that resembles a coherent state, whilst $\Delta_{\alpha}^{\mathrm{diag}}$ is diagonal in the number state basis. Noting the independence of $\Delta_{\alpha}^{\mathrm{diag}}$ from $\phi_{\alpha}$ and the independence of both contributions from $\beta$, we can write
\begin{equation}
    T_{r_{\alpha}} = \left\| \Delta_{\alpha}^{\mathrm{diag}} + \int_{\phi_{\alpha}}  P(\phi_{\alpha}|r_{\alpha}) \Delta_{\alpha}^{\mathrm{coh}} d\phi_{\alpha} \right\|,\quad
    P(\phi_{\alpha}|r_{\alpha}) = \int_{\beta} P(\phi_{\alpha},\beta|r_{\alpha}) d^2\beta.
\end{equation}
Using the triangle inequality, we bound $T_{r_{\alpha}}$ with
\begin{equation}
    T_{r_{\alpha}} \leq \left\| \Delta_{\alpha}^{\mathrm{diag}}\right\| + \left\|\int_{\phi_{\alpha}}  P(\phi_{\alpha}|r_{\alpha}) \Delta_{\alpha}^{\mathrm{coh}} d\phi_{\alpha} \right\|.
\end{equation}

Now let us assume that $(1-\lambda_y^2)^{-2}-1\geq(1-\lambda_x^2)^{2}$. If this condition holds then every term in Eq.~(\ref{eq: diag cont}) is positive for every value of $r_{\alpha}$ (if not, then the $m=0$ term may be negative). We will discuss how to adjust the bound if this is not the case later. Recalling that $\sum_m \chi_{x,m} = 1$, we evaluate
\begin{equation}
    \left\| \Delta_{\alpha}^{\mathrm{diag}}\right\| = f(x,y,r_{\alpha}) = 1 - e^{-r_{\alpha}^2 \left(1 - \tau_{x,y}\right)}g_{x,y}\Omega_{x,y}.
\end{equation}
We then evaluate
\begin{equation}
    \left\|\int_{\phi_{\alpha}}  P(\phi_{\alpha}|r_{\alpha}) \Delta_{\alpha}^{\mathrm{coh}} d\phi_{\alpha} \right\| = f(x,y,r_{\alpha}) \left\|-\int_{\phi_{\alpha}} P(\phi_{\alpha}|r_{\alpha}) \left|\sqrt{\tau_{x,y}}\alpha_{\mathrm{coh}}\middle>\middle<\sqrt{\tau_{x,y}}\alpha_{\mathrm{coh}}\right| d\phi_{\alpha} \right\|.
\end{equation}
Recalling that the term on the right-hand side that we take the norm of is (minus) our original state passed through a lossy channel, and that it therefore has a norm of $1$, we write
\begin{equation}
    \left\|\int_{\phi_{\alpha}}  P(\phi_{\alpha}|r_{\alpha}) \Delta_{\alpha}^{\mathrm{coh}} d\phi_{\alpha} \right\| = f(x,y,r_{\alpha}).
\end{equation}
Combining our results, we can write $T_{r_{\alpha}} \leq 2f(x,y,r_{\alpha})$.

We can then bound the diamond norm with
\begin{equation}
    D_{x,y,E} \leq \sup_{P(r_{\alpha})\in\mathcal{P}_E} 2\int_{r_{\alpha}} f(x,y,r_{\alpha}) P(r_{\alpha}) dr_{\alpha},\quad
    \mathcal{P}_E = \left\{P(r_{\alpha})|\int_{r_{\alpha}} r_{\alpha}^2 P(r_{\alpha}) dr_{\alpha}\leq E\right\}.
\end{equation}
Via calculus of variations, we maximise over all probability distributions, $P(r_{\alpha})$, and find that the optimal distribution is the delta function centred on $r_{\alpha}=\sqrt{E}$. We therefore finally have the bound in Eq.~(\ref{eq: lossy channel comp bound}). Due to the use of the triangle inequality, the bound is not tight.

Let us now consider cases in which $(1-\lambda_y^2)^{-2}-1<(1-\lambda_x^2)^{2}$. In this case, the first ($m=0$) term of Eq.~(\ref{eq: diag cont}) will be negative for sufficiently small $r_{\alpha}$. All of the other terms ($m\geq 1$) will still be positive. In fact, the $m=0$ term is unbounded from below: $(1-\chi_{y,0}^2)^{-\frac{1}{2}}$ can be arbitrarily large, so $\left\| \Delta_{\alpha}^{\mathrm{diag}}\right\|$ (and $\left\| \Delta_{\alpha}^{\mathrm{coh}}\right\|$) can too. This is not unphysical, since $\Delta_{\alpha}^{\mathrm{diag}}$ and $\Delta_{\alpha}^{\mathrm{coh}}$ diverge in opposite directions for small $\lambda_y$ and so cancel each other out. Rather, it is a quirk of how we have chosen to split $\Delta_{\alpha}$ into two contributions. It does mean, however, that bounding $\left\| \Delta_{\alpha} \right\|$ using $\left\| \Delta_{\alpha}^{\mathrm{diag}}\right\|$ and $\left\| \Delta_{\alpha}^{\mathrm{coh}}\right\|$ results in a bad bound.

Instead, we can split $\Delta_{\alpha}$ into three contributions:
\begin{align}
    &\Delta_{\alpha}^{\mathrm{diag}} = \chi_{x,0}\left|0\middle>\middle<0\right| + \sum_{m=1}^{\infty} \chi_{x,m}\left(1-e^{-r_{\alpha}^2 \left(1 - \tau_{x,y}\right)}g_{x,y}(1-\chi_{y,m}^2)^{-\frac{1}{2}}\right)\left|m\middle>\middle<m\right|,\\
    &\Delta_{\alpha}^{\mathrm{coh}} = -\left(1 - e^{-r_{\alpha}^2 \left(1 - \tau_{x,y}\right)}g_{x,y}\Omega'_{x,y}\right)\left|\sqrt{\tau_{x,y}}\alpha_{\mathrm{coh}}\middle>\middle<\sqrt{\tau_{x,y}}\alpha_{\mathrm{coh}}\right|,\\
    &\Delta_{\alpha}^{\mathrm{extra}} = e^{-r_{\alpha}^2 \left(1 - \tau_{x,y}\right)}g_{x,y}\chi_{x,0}(1-\chi_{y,0}^2)^{-\frac{1}{2}}\left(\left|\sqrt{\tau_{x,y}}\alpha_{\mathrm{coh}}\middle>\middle<\sqrt{\tau_{x,y}}\alpha_{\mathrm{coh}}\right| - \left|0\middle>\middle<0\right|\right),
\end{align}
where $\Omega'_{x,y}=\Omega_{x,y}-\chi_{x,0}(1-\chi_{y,0}^2)^{-\frac{1}{2}}$. Again using the triangle inequality, we get
\begin{equation}
    T_{r_{\alpha}} \leq \left\| \Delta_{\alpha}^{\mathrm{diag}}\right\| + \left\| \Delta_{\alpha}^{\mathrm{coh}} \right\| + \left\| \Delta_{\alpha}^{\mathrm{extra}} \right\|,
\end{equation}
where we now have $\left\| \Delta_{\alpha}^{\mathrm{diag}}\right\|=\left\| \Delta_{\alpha}^{\mathrm{coh}}\right\|=f'(x,y,r_{\alpha})=1 - e^{-r_{\alpha}^2 \left(1 - \tau_{x,y}\right)}g_{x,y}\Omega'_{x,y}$. Since $\left|\sqrt{\tau_{x,y}}\alpha_{\mathrm{coh}}\right>$ and $\left|0\right>$ are both pure, we can calculate the trace norm of $\Delta_{\alpha}^{\mathrm{extra}}$ exactly using the fidelity. We get
\begin{equation}
    \left\| \Delta_{\alpha}^{\mathrm{extra}} \right\| = 2e^{-r_{\alpha}^2 \left(1 - \tau_{x,y}\right)}g_{x,y}\chi_{x,0}\sqrt{\frac{1-e^{-r_{\alpha}^2\tau_{x,y}}}{1-\chi_{y,0}^2}},
\end{equation}
which allows us to bound $T_{r_{\alpha}}$ and hence the diamond norm (again using calculus of variations, although this time $T_{r_{\alpha}}$ has a more complicated dependence on $\alpha$). Note that this bound on $T_{r_{\alpha}}$ is still not necessarily tight, even if it does not diverge for small $\lambda_y$, the way the original bound did.

\section{Bounding the diamond norm from an energy-dependent replacement channel}\label{app: edrc comp}

Defining $\Delta_{\alpha}=\mathcal{E}_{x,y}[\left|\alpha_{\mathrm{coh}}\middle>\middle<\alpha_{\mathrm{coh}}\right|] - \mathcal{C}_{x,y}[\left|\alpha_{\mathrm{coh}}\middle>\middle<\alpha_{\mathrm{coh}}\right|]$, we get
\begin{equation}
    \Delta_{\alpha} = e^{-(1-\tau_{x,y})|\alpha|^2}g_{x,y} \sum_{m=0}^{\infty} \chi_{x,m}\left(\Omega_{x,y} - (1-\chi_{y,m}^2)^{-\frac{1}{2}}\right)\left|m\middle>\middle<m\right|.
\end{equation}
If we then write our input state as
\begin{equation}
    \psi_{IS} = \int_{\alpha} \int_{\beta} P(r_{\alpha})P(\phi_{\alpha},\beta|r_{\alpha}) \left|\beta_{\mathrm{coh}}\middle>\middle<\beta_{\mathrm{coh}}\right|\otimes\left|r_{\alpha} e^{i\phi_{\alpha}}_{\mathrm{coh}}\middle>\middle<r_{\alpha} e^{i\phi_{\alpha}}_{\mathrm{coh}}\right| d^2 \alpha~d^2 \beta,
\end{equation}
the trace norm between the channel outputs is
\begin{equation}
    T_{\psi} = g_{x,y} \sum_{m=0}^{\infty} \chi_{x,m}\left|\Omega_{x,y} - (1-\chi_{y,m}^2)^{-\frac{1}{2}}\right| \int P(\phi_{\alpha},\beta|r_{\alpha})d\phi_{\alpha}~d^2 \beta \int P(r_{\alpha}) e^{-(1-\tau_{x,y})r_{\alpha}^2} dr_{\alpha},
\end{equation}
which (recalling that $\Tr \Delta_{\alpha} = 0$) we can rewrite as
\begin{equation}
    T_{\psi} = 2g_{x,y} \sum_{m=0}^{m_c} \chi_{x,m}\left((1-\chi_{y,m}^2)^{-\frac{1}{2}}-\Omega_{x,y}\right)\int P(r_{\alpha}) e^{-(1-\tau_{x,y})r_{\alpha}^2} dr_{\alpha},
\end{equation}
where $m_c$ is the largest integer for which $(1-\chi_{y,m}^2)^{-\frac{1}{2}}>\Omega_{x,y}$. This is maximised by setting $P(r_{\alpha})$ to a delta function centred on $0$. Thus, we recover Eq.~(\ref{eq: edrc diamond}).

\section{$N$ port formalism}\label{app: N ports}

Here, we will develop the $N$ port formalism and find an expression for the measurement operator in terms of the eigenvectors of $\rho_y$. We find these eigenvectors up to a set of parameters, $\{\eta^{(i)}\}$, that are to be determined from an equation, and in Subsection~\ref{app: N port eigendecomposition} we show how to calculate these parameters for arbitrary $N$ (although we do not give explicit expressions for them, in the general case, in terms of $\lambda_y$). We then show how to obtain the PBT channel output for any channel input, using this parameter set. This is contingent on determining the parameter set, $\{\eta^{(i)}\}$, which can be regarded as the difficult part of the calculation, especially analytically. In Appendix~\ref{app: 3 ports}, we solve this problem analytically for the three port case, and so give a complete solution for $N=3$. In Subsection~\ref{app: N port numerics}, we show how to calculate the channel output numerically.

A more general, $N$ port formulation of the effect of $\rho_y$ on a generic state (i.e. a more general form of Eq.~(\ref{eq: rho gen effect 2 ports})) is given by
\begin{equation}
     \rho_y \left|p\right>_{C}\left|\mathcal{Q}\right>_{A} = (1-\lambda_y^2) \sum_{r=0}^{\infty}(-\lambda_y)^{p+r} \left|r\right>_{C} \otimes \sum_{i=1}^N \delta_{pq_i} \left|\Bar{\mathcal{Q}}_i\right>_{A_{\Bar{i}}}\left|r\right>_{A_i},
\end{equation}
where $\mathcal{Q}=\{q_1,q_2,...q_N\}$ is a multiset (a set with repetition allowed) of $N$ integers and $\Bar{\mathcal{Q}}_i$ is that same multiset with the $i$-th element removed. Any state that does not lie in the kernel of $\rho_y$ must therefore take the form
\begin{equation}
    \left|\psi(f)\right>_{CA} = \sum_{\mathcal{M}} \sum_{n=0}^{\infty} \left|n\right>_C \otimes \sum_{\Phi\in\mathcal{P}} f(\mathcal{M},n,\Phi) \left(\Phi\left|n\right>\left|\mathcal{M}\right>\right)_A,\label{eq: gen state N ports}
\end{equation}
where the first sum is over all distinct multisets of $N-1$ integers, $\mathcal{M}=\{m_1,...m_{N-1}\}$, $\Phi$ is a permutation of the $N$ ports (a way of reordering the multiset composed of $n$ and the $N-1$ elements of $\mathcal{M}$, some of which may be repeated elements), and $\mathcal{P}$ is the set of all such non-degenerate permutations (if $\mathcal{M}$ has repeated elements, some permutations will result in the same sequences). $f$ is a function that defines a particular state (i.e. it defines an assignment of coefficients). This is a generalisation of Eq.~(\ref{eq: gen state 2 ports}). Note that whilst Eq.~(\ref{eq: gen state N ports}) does involve summing over every possible multiset $\mathcal{M}$, for full generality, we will soon discover that each eigenvector of $\rho$ only involves a single value $\mathcal{M}$.

Applying $\rho_y$ to the state in Eq.~(\ref{eq: gen state N ports}) gives
\begin{equation}
    \begin{split}
        \rho_y\left|\psi(f)\right>_{CA} = (1-\lambda_y^2)&\sum_{\mathcal{M}} \sum_{q=0}^{\infty} (-\lambda_y)^q \left|q\right>_C\\
        &\otimes \sum_{\Phi\in\mathcal{P}} \left( \sum_{\substack{n=0,\\n\notin \mathcal{M}}}^{\infty} (-\lambda_y)^n f(\mathcal{M},n,\Phi) + \sum_{m\in \bar{\mathcal{M}}} (-\lambda_y)^{m} \sum_{\Phi'\in\mathcal{P}_m} f(\mathcal{M},m,\Phi\cdot\Phi') \right)\left(\Phi\left|q\right>\left|\mathcal{M}\right>\right)_A,
    \end{split}
\end{equation}
where $\bar{\mathcal{M}}$ is the set of unique elements of $\mathcal{M}$, $\mathcal{P}_m$ is the subset of $\mathcal{P}$ that only permutes $n$ and elements of $\mathcal{M}$ that take the value $m$ (including the identity), and $\Phi\cdot\Phi'$ denotes the composition of the two permutations. We then derive an eigenvector condition:
\begin{equation}
    \Xi (1-\lambda_y^2)^{-1} (-\lambda_y)^{-q} f(\mathcal{M},q,\Phi) = \sum_{\substack{n=0,\\n\notin \mathcal{M}}}^{\infty} (-\lambda_y)^n f(\mathcal{M},n,\Phi) + \sum_{m\in \bar{\mathcal{M}}} (-\lambda_y)^{m} \sum_{\Phi'\in\mathcal{P}_m} f(\mathcal{M},m,\Phi\cdot\Phi')\quad\forall \Phi,\forall q\notin\mathcal{M}.\label{eq: eigenvec cond q notin m}
\end{equation}
Once again, there is no $q$-dependence on the right-hand side, so we can write
\begin{equation}
    f(\mathcal{M},q\notin\mathcal{M},\Phi) = c_{\mathcal{M},\Phi}(-\lambda_y)^q.\label{eq: f q notin m}
\end{equation}
Substituting Eq.~(\ref{eq: f q notin m}) into Eq.~(\ref{eq: eigenvec cond q notin m}) (and assuming $c_{\mathcal{M},\Phi}\neq 0$; we will discuss how to deal with this case later), we get
\begin{align}
    &\Xi = g(\mathcal{M}) + h(\mathcal{M},c_{\mathcal{M}})\quad\forall \Phi,\label{eq: gen eigval}\\
    &g(\mathcal{M})=1-(1-\lambda_y^2)\sum_{m\in \bar{\mathcal{M}}}\lambda_y^{2m},
    \quad h(\mathcal{M},c_{\mathcal{M}}) = \frac{1-\lambda_y^2}{c_{\mathcal{M},\Phi}}\sum_{m\in \bar{\mathcal{M}}} (-\lambda_y)^{m} \sum_{\Phi'\in\mathcal{P}_m} f(\mathcal{M},m,\Phi\cdot\Phi'),
\end{align}
where we note that $h$ can have no $\Phi$-dependence, since the eigenvalue is fixed. The value of $h$ will be different for different eigenvectors, which correspond to different choices of variables $c_{\mathcal{M},\Phi}$ (we denote by $c_{\mathcal{M}}$ the collection of all of these variables for every value of $\Phi$) and $f(\mathcal{M},m\in\mathcal{M},\Phi)$, but any choice of these variables must make $h$ a constant with regard to $\Phi$.

The eigenvector condition for $q\in\mathcal{M}$ is
\begin{equation}
    \Xi (-\lambda_y)^{-q} \sum_{\Phi'\in\mathcal{P}_q} f(\mathcal{M},q,\Phi\cdot\Phi') = \sum_{\Phi'\in\mathcal{P}_q}\left(g(\mathcal{M}) c_{\mathcal{M},\Phi\cdot\Phi'} + (1-\lambda_y^2)\sum_{m\in \bar{\mathcal{M}}} (-\lambda_y)^{m} \sum_{\Phi''\in\mathcal{P}_m} f(\mathcal{M},m,\Phi\cdot\Phi'\cdot\Phi'')\right),
\end{equation}
and using the $\Phi$-independence of $h$, we get
\begin{equation}
    \Xi (-\lambda_y)^{-q} \sum_{\Phi'\in\mathcal{P}_q} f(\mathcal{M},q,\Phi\cdot\Phi') = \left(g(\mathcal{M}) + h(\mathcal{M},c_{\mathcal{M}})\right)\sum_{\Phi'\in\mathcal{P}_q} c_{\mathcal{M},\Phi\cdot\Phi'}.
\end{equation}
Finally, we can choose without loss of generality that $\Xi (-\lambda_y)^{-q} f(\mathcal{M},q,\Phi\cdot\Phi') = \left(g(\mathcal{M}) + h(\mathcal{M},c_{\mathcal{M}})\right) c_{\mathcal{M},\Phi\cdot\Phi'}$, and so
\begin{equation}
    f(\mathcal{M},q,\Phi) = c_{\mathcal{M},\Phi}(-\lambda_y)^q.\label{eq: f all q}
\end{equation}
The only difference between Eqs.~(\ref{eq: f q notin m}) and (\ref{eq: f all q}) is that the expression now applies to all values of $q$, rather than just those that are not elements of $\mathcal{M}$.

This tells us that all eigenvectors of $\rho_y$ can be expressed as (Eq.~(\ref{eq: eigenstate gen}) in the main text)
\begin{equation*}
    \left|\psi(\mathcal{M},i)\right>_{CA} = \sqrt{\frac{1-\lambda_y^2}{\Xi^{(i)}_{\mathcal{M}}}}\sum_{n=0}^{\infty} (-\lambda_y)^n\left|n\right>_C \otimes \sum_{\Phi\in\mathcal{P}} \eta^{(i)}_{\mathcal{M},\Phi} \left(\Phi\left|n\right>\left|\mathcal{M}\right>\right)_A,
\end{equation*}
where we no longer sum over different choices of $\mathcal{M}$, since any states of the form in Eq.~(\ref{eq: eigenstate gen}) with different values of $\mathcal{M}$ are orthogonal, and where we have made the replacement $\eta_{\mathcal{M}}=\sqrt{\frac{1-\lambda_y^2}{\Xi_{\mathcal{M},\eta_{\mathcal{M}}}}}c_{\mathcal{M}}$ for easier normalisation. We rewrite $h(\mathcal{M},\eta_{\mathcal{M}})$ as
\begin{equation}
    h(\mathcal{M},\eta^{(i)}_{\mathcal{M}}) = \frac{1-\lambda_y^2}{\eta^{(i)}_{\mathcal{M},\Phi}}\sum_{m\in \bar{\mathcal{M}}} \lambda_y^{2m} \sum_{\Phi'\in\mathcal{P}_m} \eta^{(i)}_{\mathcal{M},\Phi\cdot\Phi'}\label{eq: gen eig cond}
\end{equation}
and use its $\Phi$-independence to find the values of the coefficients $\eta_{\mathcal{M}}$ that give valid eigenvectors. The normalisation condition for eigenvectors of the form given in Eq.~(\ref{eq: eigenstate gen}) is $\sum_{\Phi\in\mathcal{P}} |\eta^{(i)}_{\mathcal{M},\Phi}|^2=1$, whilst orthogonality demands that for any pair of orthogonal eigenvectors, $\left|\psi(\mathcal{M},\eta^{(\alpha)}_{\mathcal{M}})\right>$ and $\left|\psi(\mathcal{M},\eta^{(\beta)}_{\mathcal{M}})\right>$, $\sum_{\Phi\in\mathcal{P}} \eta^{(\alpha)}_{\mathcal{M},\Phi}\eta^{(\beta)*}_{\mathcal{M},\Phi}=0$. From this orthonormality condition, it is clear that the total number of independent eigenvectors for a given multiset $\mathcal{M}$ is at most $|\mathcal{P}|$. We will address how these eigenvectors can be found in Subsection~\ref{app: N port eigendecomposition}. There is a subtlety here: if $\eta_{\mathcal{M},\Phi}=0$ for some particular $\Phi$, we instead require that $\sum_{m\in \bar{\mathcal{M}}} \lambda_y^{2m} \sum_{\Phi'\in\mathcal{P}_m}\eta_{\mathcal{M},\Phi\cdot\Phi'}=0$.

Now consider the general form of $\sigma_{1,y}$ applied to a generic state (generalising Eq.~(\ref{eq: 2 port sigma on gen})):
\begin{equation}
     \sigma_{1,y} \left|p\right>_{C}\left|\mathcal{Q}\right>_{A} = \delta_{pq_1}(1-\lambda_y^2) \sum_{r=0}^{\infty}(-\lambda_y)^{p+r} \left|r\right>_{C}\left|r\right>_{A_1}\left|\Bar{\mathcal{Q}}_1\right>_{A_{\Bar{1}}}.
\end{equation}
Define $\Tilde{\mathcal{P}}$ as the subset of $\mathcal{P}$ containing all permutations that leave the first element unchanged. Observe that any element in $\mathcal{P}$ can be uniquely written as a composition of an element of $\mathcal{P}_m$ (for some $m$) and an element of $\Tilde{\mathcal{P}}$. Then, applying $\sigma_{1,y}^{[N]}$ to an eigenstate of $\rho_y$ gives
\begin{equation}
    \sigma_{1,y}\left|\psi(\mathcal{M},i)\right> = \frac{(1-\lambda_y^2)^{\frac{3}{2}}}{\Xi_{\mathcal{M}}^{(i)\frac{1}{2}}}\sum_{r=0}^{\infty} (-\lambda_y)^{r}\left|r\right>\otimes \sum_{\Phi\in\Tilde{\mathcal{P}}}\bigg[ \sum_{\substack{n=0,\\n\notin \mathcal{M}}}^{\infty} \lambda_y^{2n} \eta^{(i)}_{\mathcal{M},\Phi} + \sum_{m\in\bar{\mathcal{M}}} \lambda_y^{2m} \sum_{\Phi\in\mathcal{P}_m}\eta^{(i)}_{\mathcal{M},\Phi\cdot\Phi'}\bigg](\Phi \left|r\right>\left|\mathcal{M}\right>).
\end{equation}
Noting the similarity of the square bracketed term to the expression for eigenvalues, we simplify this to
\begin{equation}
    \sigma_{1,y}\left|\psi(\mathcal{M},i)\right> = \sqrt{(1-\lambda_y^2)\Xi^{(i)}_{\mathcal{M}}}\sum_{\Phi\in\Tilde{\mathcal{P}}}\eta^{(i)}_{\mathcal{M},\Phi} \sum_{r=0}^{\infty} (-\lambda_y)^{r}\left|r\right>(\Phi \left|r\right>\left|\mathcal{M}\right>),
\end{equation}
and then calculate
\begin{equation}
    \begin{split}
        \left<\psi(\mathcal{M},\alpha)\right| \sigma_{1,y} \left|\psi(\mathcal{M},\beta)\right> &= (1-\lambda_y^2)\sqrt{\frac{\Xi_{\mathcal{M}}^{(\beta)}}{\Xi_{\mathcal{M}}^{(\alpha)}}} \sum_{\Phi\in\Tilde{\mathcal{P}}} \eta_{\mathcal{M},\Phi}^{(\beta)}\bigg[ \sum_{\substack{n=0,\\n\notin \mathcal{M}}}^{\infty} \lambda_y^{2n} \eta_{\mathcal{M},\Phi}^{(\alpha)*} + \sum_{m\in\bar{\mathcal{M}}} \lambda_y^{2m} \sum_{\Phi\in\mathcal{P}_m}\eta_{\mathcal{M},\Phi\cdot\Phi'}^{(\alpha)}\bigg]\\
        &= \sqrt{\Xi_{\mathcal{M}}^{(\alpha)}\Xi_{\mathcal{M}}^{(\beta)}}\sum_{\Phi\in\Tilde{\mathcal{P}}}\eta_{\mathcal{M},\Phi}^{(\alpha)*}\eta_{\mathcal{M},\Phi}^{(\beta)}.
    \end{split}
\end{equation}
Consequently, we can write
\begin{equation}
    \rho_y^{-\frac{1}{2}}\sigma_{1,y}\rho_y^{-\frac{1}{2}}= \sum_{\mathcal{M}} \sum_{\eta_{\mathcal{M}}^{(\alpha)},\eta_{\mathcal{M}}^{(\beta)}} \Big( \sum_{\Phi\in\Tilde{\mathcal{P}}}\eta_{\mathcal{M},\Phi}^{(\alpha)*}\eta_{\mathcal{M},\Phi}^{(\beta)} \Big) \left|\psi(\mathcal{M},\alpha)\middle>\middle<\psi(\mathcal{M},\beta)\right|,
\end{equation}
where the sum is over those values of $\eta_{\mathcal{M}}^{(\alpha)}$ and $\eta_{\mathcal{M}}^{(\beta)}$ that give rise to an orthonormal set of eigenvectors. Finally, we arrive at Eq.~(\ref{eq: measurement N modes}), which gives us an explicit expression for the measurement operator so long as we have the eigendecomposition of $\rho_y$.

Numerically, Eq.~(\ref{eq: measurement N modes}) is sufficient to calculate the channel output for any input, with an appropriate truncation on all systems and a maximum value of $\mathcal{M}$. Nonetheless, we can reduce the complexity of the calculation significantly with a little more work.

We define $\phi^{(1)}_{ab,x}$, similarly to in Eq.~(\ref{eq: arbitrary number state}), as
\begin{equation}
    \begin{split}
        \phi^{(1)}_{ab,x} &= (1-\lambda_x^2) \left|a\middle>\middle<b\right|_C \otimes \left( \sum_{p,q=0}^{\infty}(-\lambda_x)^{p+q} \left|pp\middle>\middle<qq\right|_{A_1 B_1}\right) \bigotimes_{i=1}^{N-1} \left( \sum_{r_i=0}^{\infty} \chi_{x,r_i} \left|r_i\middle>\middle<r_i\right|_{A_i} \right)\\
        &= (1-\lambda_x^2)^N \left|a\middle>\middle<b\right|_C \otimes \sum_{p,q,\{r_i\}=0}^{\infty}(-\lambda_x)^{p+q+2\sum r_i} \left|p r_1 \ldots r_{N-1} p\middle>\middle<q r_1 \ldots r_{N-1} q\right|_{A B_1},
    \end{split}\label{eq: arbitrary number state N modes}
\end{equation}
where the subscript on the second sum indicates that $p$, $q$, and each of the $N-1$ indices $r_i$ are summed from $0$ to $\infty$. By calculating $\mathcal{E}_{x,y}[\left|a\middle>\middle<b\right|] = N \Tr_{AC}[M^{(1)}_{y} \phi^{(1)}_{ab,x}]$, we obtain an expression for the effect of the teleportation channel on an arbitrary component of the number state basis, and hence can find the output state for any given input state. Using Eq.~(\ref{eq: measurement N modes}), we get
\begin{equation}
    \mathcal{E}_{x,y}[\left|a\middle>\middle<b\right|] = \delta_{ab}(1-\lambda_x^2)\sum_{n=0}^{\infty} \lambda_x^{2n} \left|n\middle>\middle<n\right|
    + N\sum_{\mathcal{M}} \sum_{\eta_{\mathcal{M}}^{(\alpha)},\eta_{\mathcal{M}}^{(\beta)}} (1-\delta_{\alpha\beta}) \Big( \sum_{\Phi\in\Tilde{\mathcal{P}}}\eta_{\mathcal{M},\Phi}^{(\alpha)*}\eta_{\mathcal{M},\Phi}^{(\beta)} \Big) 
    \left<\psi(\mathcal{M},\beta)\right| \phi^{(1)}_{ab,x} \left|\psi(\mathcal{M},\alpha)\right>,
\end{equation}
where the first term comes from the identity term in Eq.~(\ref{eq: measurement N modes}). Applying Eq.~(\ref{eq: arbitrary number state N modes}), the summed over terms become
\begin{equation}
    \begin{split}
        \left<\psi(\mathcal{M},\beta)\right| \phi^{(1)}_{ab,x} \left|\psi(\mathcal{M},\alpha)\right> =
        \frac{1-\lambda_y^2}{\sqrt{\Xi_{\mathcal{M}}^{(\alpha)}\Xi_{\mathcal{M}}^{(\beta)}}}(1-\lambda_x^2)^{N}(-\lambda_y)^{a+b}
        \sum_{p,q,\{r_i\}=0}^{\infty}(-\lambda_x)^{p+q+2\sum r_i} \sum_{\Phi',\Phi''\in\mathcal{P}} \eta_{\mathcal{M},\Phi'}^{(\alpha)}\eta_{\mathcal{M},\Phi''}^{(\beta)*}\times\\
        \left<q r_1 \ldots r_{N-1}\right|\Phi'\left(\left|b\right>\left|\mathcal{M}\right>\right)\left(\left<p r_1 \ldots r_{N-1}\right|\Phi''\left(\left|a\right>\left|\mathcal{M}\right>\right)\right)^*\left|p\middle>\middle<q\right|.
    \end{split}\label{eq: alpha beta a b contraction}
\end{equation}
Next, we carry out the sum over $\alpha$ and $\beta$ (i.e. over the different eigenvectors corresponding to the same multiset $\mathcal{M}$. Per Eq.~(\ref{eq: gamma expr}) in the main text, we define
\begin{equation*}
    \Gamma(\mathcal{M}) = \sum_{\eta_{\mathcal{M}}^{(\alpha)},\eta_{\mathcal{M}}^{(\beta)}} \frac{1-\delta_{\alpha\beta}}{\sqrt{\Xi_{\mathcal{M}}^{(\alpha)}\Xi_{\mathcal{M}}^{(\beta)}}} \Big( \sum_{\Phi\in\Tilde{\mathcal{P}}}\eta_{\mathcal{M},\Phi}^{(\alpha)*}\eta_{\mathcal{M},\Phi}^{(\beta)} \Big) \eta_{\mathcal{M}}^{(\beta)\dagger}\eta_{\mathcal{M}}^{(\alpha)},\quad
    \Gamma(\mathcal{M},\Phi',\Phi'') = \Gamma(\mathcal{M})_{\Phi',\Phi''},
\end{equation*}
Recall that the subscripted $\Phi$ denotes a specific element of the vector $\eta_{\mathcal{M}}$ but without having to specify a particular basis for the vector (since this would require us to define some general scheme for numbering the permutations). Hence, $\Gamma(\mathcal{M},\Phi',\Phi'')$ refers to a particular column (indexed by $\Phi'$) and row (indexed by $\Phi''$) of the matrix $\Gamma(\mathcal{M})$. Note that $\Gamma(\mathcal{M})$ has $\lambda_y$-dependence, because the vectors $\eta_{\mathcal{M}}$ also depend on $\lambda_y$. Now, we can write
\begin{equation}
    \begin{split}
        \mathcal{E}_{x,y}[\left|a\middle>\middle<b\right|] = \delta_{ab}(1-\lambda_x^2)&\sum_{n=0}^{\infty} \lambda_x^{2n} \left|n\middle>\middle<n\right|
        + N (1-\lambda_x^2)^{N}(1-\lambda_y^2)(-\lambda_y)^{a+b} \sum_{\mathcal{M}}
        \sum_{p,q,\{r_i\}=0}^{\infty}(-\lambda_x)^{p+q+2\sum r_i}\times\\
        &\sum_{\Phi',\Phi''\in\mathcal{P}} \Gamma(\mathcal{M},\Phi',\Phi'')
        \left<q r_1 \ldots r_{N-1}\right|\Phi'\left(\left|b\right>\left|\mathcal{M}\right>\right)\left(\left<p r_1 \ldots r_{N-1}\right|\Phi''\left(\left|a\right>\left|\mathcal{M}\right>\right)\right)^*\left|p\middle>\middle<q\right|.
    \end{split}\label{eq: N mode ab channel}
\end{equation}

To understand what this means, let us examine the term $\left<p r_1 \ldots r_{N-1}\right|\Phi''\left(\left|a\right>\left|\mathcal{M}\right>\right)$. This term is equal to $1$ if the sequence $(p,r_1,\ldots,r_{N-1})$ is the same as the sequence $\Phi''(a,\mathcal{M})$, where $\Phi''$ denotes some specific reordering of the sequence on which it acts. Recall that we sum over distinct multisets $\mathcal{M}$, but over every choice of $r_1$, $r_2$, etc., so that the elements of the sequence $(p,r_1,\ldots,r_{N-1})$ will form the same multiset for different choices of $\{r_i\}$. E.g. $p=1$, $r_1=3$, $r_2=4$ forms a different sequence $(p,r_1,r_2)$ than $p=1$, $r_1=4$, $r_2=3$. Each term $\left<q r_1 \ldots r_{N-1}\right|\Phi'\left(\left|b\right>\left|\mathcal{M}\right>\right)$ can only be non-zero if the multisets $\{q,r_i\}$ and $\{b,\mathcal{M}\}$ are the same (have all of the same elements with the same multiplicities) and similarly $\left<p r_1 \ldots r_{N-1}\right|\Phi''\left(\left|a\right>\left|\mathcal{M}\right>\right)$ is can only be non-zero if the multisets $\{p,r_i\}$ and $\{a,\mathcal{M}\}$ are the same (although they will still only be non-zero for specfic choices of $\Phi$). These two conditions can only be satisfied simultaneously if $a=b$ and $p=q$ or if $a\neq b$, $p=a$, and $q=b$.

If $a \neq b$, Eq.~(\ref{eq: N mode ab channel}) becomes
\begin{equation}
    \begin{split}
        \mathcal{E}_{x,y}[\left|a\middle>\middle<b\right|] = &N (1-\lambda_x^2)^{N}(1-\lambda_y^2)(\lambda_x \lambda_y)^{a+b} \sum_{\mathcal{M}}
        \sum_{\{r_i\}=0}^{\infty} \lambda_x^{2\sum_{m\in \mathcal{M}} m}\times\\
        &\sum_{\Phi',\Phi''\in\mathcal{P}} \Gamma(\mathcal{M},\Phi',\Phi'')
        \left<b r_1 \ldots r_{N-1}\right|\Phi'\left(\left|b\right>\left|\mathcal{M}\right>\right)\left<a r_1 \ldots r_{N-1}\right|\Phi''\left(\left|a\right>\left|\mathcal{M}\right>\right)\left|a\middle>\middle<b\right|,
    \end{split}\label{eq: N mode ab channel a neq b}
\end{equation}
whilst if $a=b$,
\begin{equation}
    \begin{split}
        \mathcal{E}_{x,y}[\left|a\middle>\middle<a\right|] = (1-\lambda_x^2)\sum_{n=0}^{\infty}& \lambda_x^{2n} \left|n\middle>\middle<n\right|
        + N (1-\lambda_x^2)^{N}(1-\lambda_y^2) (\lambda_x\lambda_y)^{2a} \sum_{\mathcal{M}}
        \sum_{n,\{r_i\}=0}^{\infty} \lambda_x^{2\sum_{m\in \mathcal{M}} m}\times\\
        &\sum_{\Phi',\Phi''\in\mathcal{P}} \Gamma(\mathcal{M},\Phi',\Phi'')
        \left<n r_1 \ldots r_{N-1}\right|\Phi'\left(\left|a\right>\left|\mathcal{M}\right>\right)\left<n r_1 \ldots r_{N-1}\right|\Phi''\left(\left|a\right>\left|\mathcal{M}\right>\right)\left|n\middle>\middle<n\right|.
    \end{split}\label{eq: N mode ab channel a=b}
\end{equation}
This tells us that $\mathcal{E}_{x,y}[\left|a\middle>\middle<a\right|]$ only has on-diagonal components, whilst $\mathcal{E}_{x,y}[\left|a\middle>\middle<b\right|]$ only has a single off-diagonal component (namely $\left|a\middle>\middle<b\right|$).

Considering the term $\left<b r_1 \ldots r_{N-1}\right|\Phi'\left(\left|b\right>\left|\mathcal{M}\right>\right)\left<a r_1 \ldots r_{N-1}\right|\Phi''\left(\left|a\right>\left|\mathcal{M}\right>\right)$ (from Eq.~(\ref{eq: N mode ab channel a neq b})), we have four different situations. If $a,b\notin \mathcal{M}$, this term is only non-zero if $\{r_i\}$ has the same elements and multiplicities as $\mathcal{M}$ and $\Phi',\Phi''\in\tilde{\mathcal{P}}$. If $a\in \mathcal{M}$ but $b\notin \mathcal{M}$, we still require that $\{r_i\}$ has the same elements and multiplicities as $\mathcal{M}$, but $\Phi''$ can now be constructed as $\Phi'\cdot\Phi_0$, where again $\Phi'\in\tilde{\mathcal{P}}$ and $\Phi_0\in\mathcal{P}_a$ (recalling that $\mathcal{P}_a$ is the set of permutations comprising the identity and swaps between the first element and any element of the sequence with value $a$). We have a similar result if $b\in \mathcal{M}$ but $a\notin \mathcal{M}$, but with $\Phi'$ now constructed as $\Phi''\cdot\Phi_0$ (where $\Phi''\in\tilde{\mathcal{P}}$ and $\Phi_0\in\mathcal{P}_b$). Finally, if $a,b\in \mathcal{M}$, we write $\Phi'=\Phi\cdot\Phi_{0}'$ and $\Phi''=\Phi\cdot\Phi_{0}''$, where $\Phi\in\tilde{\mathcal{P}}$, $\Phi_{0}''\in\mathcal{P}_a$, and $\Phi_{0}'\in\mathcal{P}_b$. Eq.~(\ref{eq: N mode ab channel a neq b}) therefore reduces to
\begin{equation}
    \begin{split}
        \mathcal{E}_{x,y}&[\left|a\middle>\middle<b\right|] = 
        N (1-\lambda_x^2)^{N}(1-\lambda_y^2)(\lambda_x \lambda_y)^{a+b}
        \Bigg(\sum_{\mathcal{M}:a,b\notin \mathcal{M}} \lambda_x^{2\sum_{m\in \mathcal{M}} m}
        \sum_{\Phi\in\tilde{\mathcal{P}}} \Gamma(\mathcal{M},\Phi,\Phi) +\\
        &\sum_{\mathcal{M}:a\in\mathcal{M},b\notin \mathcal{M}} \lambda_x^{2\sum_{m\in \mathcal{M}} m}
        \sum_{\Phi\in\tilde{\mathcal{P}},\Phi'\in\mathcal{P}_a} \Gamma(\mathcal{M},\Phi,\Phi\cdot\Phi') +
        \sum_{\mathcal{M}:a\notin\mathcal{M},b\in \mathcal{M}} \lambda_x^{2\sum_{m\in \mathcal{M}} m}
        \sum_{\Phi\in\tilde{\mathcal{P}},\Phi'\in\mathcal{P}_b} \Gamma(\mathcal{M},\Phi\cdot\Phi',\Phi) +\\
        &\sum_{\mathcal{M}:a,b\in \mathcal{M}} \lambda_x^{2\sum_{m\in \mathcal{M}} m}
        \sum_{\Phi\in\tilde{\mathcal{P}},\Phi'\in\mathcal{P}_b,\Phi''\in\mathcal{P}_a} \Gamma(\mathcal{M},\Phi\cdot\Phi',\Phi\cdot\Phi'')\Bigg) \left|a\middle>\middle<b\right|.
    \end{split}\label{eq: a neq b gen expr N modes}
\end{equation}

Now considering the term $\left<n r_1 \ldots r_{N-1}\right|\Phi'\left(\left|a\right>\left|\mathcal{M}\right>\right)\left<n r_1 \ldots r_{N-1}\right|\Phi''\left(\left|a\right>\left|\mathcal{M}\right>\right)$ (from Eq.~(\ref{eq: N mode ab channel a=b})), we again have four situations. If $a=n$ and $a\notin \mathcal{M}$, we require that $\Phi'=\Phi''\in\tilde{\mathcal{P}}$. If $a=n$ and $a\in \mathcal{M}$, we can write $\Phi'=\Phi\cdot\Phi_{0}'$ and $\Phi''=\Phi\cdot\Phi_{0}''$, where $\Phi\in\tilde{\mathcal{P}}$ and $\Phi_{0}',\Phi_{0}''\in\mathcal{P}_a$. If $a\neq n$ and $a\notin \mathcal{M}$, we require $n\in \mathcal{M}$ and $\Phi'=\Phi''=\Phi\cdot\Phi_0$, where $\Phi\in\tilde{\mathcal{P}}$, $\Phi_0\in\mathcal{P}_n$, and $\Phi_0\neq\mathcal{I}$. Finally, if $a\neq n$ and $a\in \mathcal{M}$, we again require $n\in \mathcal{M}$, and then can set $\Phi'=\Phi_{2}\cdot\Phi_{1}$ and $\Phi''=\Phi'\cdot\Phi_{0}$, where $\Phi_{2}\in\tilde{\mathcal{P}}$, $\Phi_1\in\mathcal{P}_n$, $\Phi_1\neq\mathcal{I}$, and $\Phi_0\in\mathcal{P}_a$. Eq.~(\ref{eq: N mode ab channel a=b}) reduces to
\begin{equation}
    \begin{split}
        \mathcal{E}_{x,y}[\left|a\middle>\middle<a\right|] = (1-\lambda_x^2)\sum_{n=0}^{\infty} \lambda_x^{2n} \left|n\middle>\middle<n\right|
        + N (1-\lambda_x^2)^{N}(1-\lambda_y^2) (\lambda_x\lambda_y)^{2a} \Bigg(\bigg( \sum_{\mathcal{M}:a\notin\mathcal{M}}
        \lambda_x^{2\sum_{m\in \mathcal{M}} m} \sum_{\Phi\in\tilde{\mathcal{P}}} \Gamma(\mathcal{M},\Phi,\Phi) +\\
        \sum_{\mathcal{M}:a\in\mathcal{M}}
        \lambda_x^{2\sum_{m\in \mathcal{M}} m} \sum_{\Phi\in\tilde{\mathcal{P}},\Phi',\Phi''\in\mathcal{P}_a} \Gamma(\mathcal{M},\Phi\cdot\Phi',\Phi\cdot\Phi'') \bigg) \left|a\middle>\middle<a\right| +\\
        \sum_{\mathcal{M}:a\notin\mathcal{M}}
        \lambda_x^{2\sum_{m\in \mathcal{M}} m} \sum_{\substack{n\in\mathcal{M}}} \sum_{\substack{\Phi\in\tilde{\mathcal{P}},\Phi'\in\mathcal{P}_n,\\ \Phi'\neq\mathcal{I}}} \Gamma(\mathcal{M},\Phi\cdot\Phi',\Phi\cdot\Phi') \left|n\middle>\middle<n\right| +\\
        \sum_{\mathcal{M}:a\in\mathcal{M}} \lambda_x^{2\sum_{m\in \mathcal{M}} m}
        \sum_{\substack{n\in\mathcal{M},\\n\neq a}} \sum_{\substack{\Phi\in\tilde{\mathcal{P}},\Phi'\in\mathcal{P}_n,\\ \Phi''\in\mathcal{P}_a,\Phi'\neq\mathcal{I}}} \Gamma(\mathcal{M},\Phi\cdot\Phi',\Phi\cdot\Phi'\cdot\Phi'') \left|n\middle>\middle<n\right| \Bigg).
    \end{split}\label{eq: a=b gen expr N modes}
\end{equation}

Eqs.~(\ref{eq: a neq b gen expr N modes}) and (\ref{eq: a=b gen expr N modes}) fully describe the $N$ port PBT channel. To use them, we only need to calculate the matrix $\Gamma(\mathcal{M})$ for each value of $\mathcal{M}$, using Eq.~(\ref{eq: gamma expr}). This requires the eigenvectors and eigenvalues for each multiset $\mathcal{M}$, which can be found using orthogonality and the condition in Eq.~(\ref{eq: gen eig cond}). We show how to do this next.

\subsection{Finding the eigenvectors of $\rho_y$}\label{app: N port eigendecomposition}

To use Eqs.~(\ref{eq: a neq b gen expr N modes}) and (\ref{eq: a=b gen expr N modes}), we need to find all of the eigenvectors of $\rho_y$ for every multiset $\mathcal{M}$. Since there are infinite multisets $\mathcal{M}$, and so infinite eigenvectors, we cannot find them individually. Instead, we categorise the multisets by their multiplicities. The eigenvectors for multisets within a given category all have a fixed form. For instance, for $N=4$, $\mathcal{M}$ can have the following forms: $\{m_1,m_1,m_1\}$ (all elements the same), $\{m_1,m_1,m_2\}$ (two unique elements), or $\{m_1,m_2,m_3\}$ (all elements unique). For a given $N$, the number of different categories of $\mathcal{M}$ that we need to consider is the number of ways of partitioning $N-1$ (we could also label the categories with the Young diagrams consisting of $N-1$ boxes).

For a fixed multiset $\mathcal{M}$, we find all sequences of $|\mathcal{P}|$ parameters $\eta_{\mathcal{M}}$ that satisfy Eq.~(\ref{eq: gen eig cond}). For a sequence of parameters, $\eta_{\mathcal{M}}$, to give rise to an eigenvector, every $\eta_{\mathcal{M},\Phi}$ in $\eta_{\mathcal{M}}$, the expression in Eq.~(\ref{eq: gen eig cond}) must take a constant value (which is not known a priori and so must also be determined).

To make the explanation of how to find the eigenvectors clearer, we will pick the specific multiset $\mathcal{M}=\{m_1,m_1,m_2\}$ (from the $4$ port case) as an example, but none of what we do is specific to this choice. Each eigenvector takes the form
\begin{equation}
    \begin{split}
        \left|\psi(\mathcal{M},\eta)\right>_{CA} = \sqrt{\frac{1-\lambda_y^2}{\Xi_{\mathcal{M},\eta}}}\sum_{n=0}^{\infty} &(-\lambda_y)^n \left|n\right>_C \otimes \big( \eta_{1}\left|n m_1 m_1 m_2\right> + \eta_{2}\left|m_2 n m_1 m_1\right> + \eta_{3}\left|m_1 m_2 n m_1\right> + \eta_{4}\left|m_1 m_1 m_2 n\right>\\
        &+ \eta_{5}\left|n m_1 m_2 m_1\right> + \eta_{6}\left|m_1 n m_1 m_2\right> + \eta_{7}\left|m_2 m_1 n m_1\right> + \eta_{8}\left|m_1 m_2 m_1 n\right>\\
        &+ \eta_{9}\left|n m_2 m_1 m_1\right> + \eta_{10}\left|m_1 n m_2 m_1\right> + \eta_{11}\left|m_1 m_1 n m_2\right> + \eta_{12}\left|m_2 m_1 m_1 n\right> \big)_A,
    \end{split}\label{eq: example eigenvector}
\end{equation}
where $\Xi_{\mathcal{M},\eta}$ is the corresponding eigenvalue. We have assigned numerical labels to the different orderings; the choice of number for each ordering is somewhat arbitrary. Previously, we had used $\Phi$ to label each ordering in a more generic way. Any of the $\eta_i$ can be complex.

Since the sequences $\eta_{\mathcal{M}}$ specify particular eigenvectors, we can represent the eigenvector $\left|\psi(\mathcal{M},\eta_{\mathcal{M}})\right>$ by the vector $\bar{\eta}$ whose elements are given by $\eta_{\mathcal{M}}$. In our example, the $i$-th eigenvector for the multiset $\mathcal{M}$ is represented by the vector $\bar{\eta}^{(i)}=(\eta^{(i)}_{1},\eta^{(i)}_{2},\eta^{(i)}_{3},\eta^{(i)}_{4},\eta^{(i)}_{5},\eta^{(i)}_{6},\eta^{(i)}_{7},\eta^{(i)}_{8},\eta^{(i)}_{9},\eta^{(i)}_{10},\eta^{(i)}_{11},\eta^{(i)}_{12})$, but we note that the actual form of the eigenvector is given by Eq.~(\ref{eq: example eigenvector}); this vector is simply a more convenient and compact representation. In this representation, orthonormality demands that $\bar{\eta}^{(i)}\cdot\bar{\eta}^{(j)}=\delta_{ij}$ for any pair of eigenvectors. Our task is then to find an orthonormal set of basis vectors, all of which obey the eigenvector condition. In our example case, each vector, $\bar{\eta}^{(i)}$, has $12$ components, and so we can find $12$ mutually orthogonal basis vectors (and hence $12$ orthogonal eigenvectors per multiset of the form $\{m_1,m_1,m_2\}$). In general, the vectors $\bar{\eta}^{(i)}$ have $|\mathcal{P}|$ components, and we can find $|\mathcal{P}|$ of them, where $|\mathcal{P}|$ is the number of ways of uniquely arranging $n$ and the elements of $\mathcal{M}$ in a sequence. If $\mathcal{M}$ has no repeated elements, this number is $N!$. The eigenvector condition then takes the form of $|\mathcal{P}|$ expressions that must all be satisfied simultaneously. The expressions overlap in terms of the variables they contain (i.e. more than one of the expressions uses each of the $\eta_j$), so it is not simple to find an assignment satisfying this condition.

Since we find that several of these eigenvectors are degenerate for the $N>2$ case, there is no unique way of finding the eigendecomposition. Instead, we present one way of finding the eigenvectors. Recall that we must ensure that
\begin{equation*}
    h(\bar{\eta}) = \frac{1-\lambda_y^2}{\eta_{\Phi}}\sum_{m\in \bar{\mathcal{M}}} \lambda_y^{2m} \sum_{\Phi'\in\mathcal{P}_m} \eta_{\Phi\cdot\Phi'}
\end{equation*}
takes a constant value for every $\Phi$. Defining $R_j$ as a rotation of all $N$ elements in a sequence by $j$ positions, we require that
\begin{equation}
    \frac{1-\lambda_y^2}{\eta_{\Phi}}\sum_{m\in \bar{\mathcal{M}}} \lambda_y^{2m} \sum_{\Phi'\in\mathcal{P}_m} \eta_{\Phi\cdot\Phi'} = \frac{1-\lambda_y^2}{\eta_{R_j\cdot\Phi}}\sum_{m\in \bar{\mathcal{M}}} \lambda_y^{2m} \sum_{\Phi'\in\mathcal{P}_m} \eta_{R_j\cdot\Phi\cdot\Phi'}.
\end{equation}
We can fulfil this requirement by setting a constant phase relation between every pair of elements $\eta_{R_j\cdot\Phi}$ and $\eta_{\Phi}$. We set
\begin{equation}
    \eta_{R_j\cdot\Phi} = e^{i\frac{2\pi jk}{N}}\eta_{\Phi},\label{eq: phase relation}
\end{equation}
for every $\Phi$ and $j$, where $k$ is an integer between $0$ and $N-1$.

Note that every permutation $\Phi$ can be uniquely decomposed as $\Phi=R_j\cdot\Phi'$, where $j$ is an integer between $0$ and $N-1$ and $\Phi'\in\Tilde{\mathcal{P}}$ (the set of permutations of only the last $N-1$ elements of the sequence, leaving the first element unchanged). Hence, by setting the phase relation between $\eta_{R_j\cdot\Phi}$ and $\eta_{\Phi}$, we divide the number of free parameters in $\Bar{\eta}$ that we need to set by $N$. Returning to our example, every eigenvector has one of the following forms:
\begin{equation}
    \begin{split}
        &\bar{\eta}^{k=0} = (\phi_1,\phi_1,\phi_1,\phi_1,\phi_2,\phi_2,\phi_2,\phi_2,\phi_3,\phi_3,\phi_3,\phi_3),\\
        &\bar{\eta}^{k=1} = (\phi_1,I\phi_1,-\phi_1,-I\phi_1,\phi_2,I\phi_2,-\phi_2,-I\phi_2,\phi_3,I\phi_3,-\phi_3,-I\phi_3),\\
        &\bar{\eta}^{k=2} = (\phi_1,-\phi_1,\phi_1,-\phi_1,\phi_2,-\phi_2,\phi_2,-\phi_2,\phi_3,-\phi_3,\phi_3,-\phi_3),\\
        &\bar{\eta}^{k=3} = (\phi_1,-I\phi_1,-\phi_1,I\phi_1,\phi_2,-I\phi_2,-\phi_2,I\phi_2,\phi_3,-I\phi_3,-\phi_3,I\phi_3),
    \end{split}
\end{equation}
where $\phi_1$, $\phi_2$, and $\phi_3$ are parameters to be determined. The eigenvector condition in all four cases is
\begin{equation}
    \frac{1-\lambda_y^2}{\eta_{1}} \big( \lambda_y^{2m_1}(\eta_{6}+\eta_{11}) + \lambda_y^{2m_2}\eta_{12} \big) =
    \frac{1-\lambda_y^2}{\eta_{5}} \big( \lambda_y^{2m_1}(\eta_{10}+\eta_{4}) + \lambda_y^{2m_2}\eta_{7} \big) =
    \frac{1-\lambda_y^2}{\eta_{9}} \big( \lambda_y^{2m_1}(\eta_{3}+\eta_{8}) + \lambda_y^{2m_2}\eta_{2} \big),\label{eq: eigenvector cond llm}
\end{equation}
except for when one of the parameters is equal to $0$. For instance, for the $k=0$ case, Eq.~(\ref{eq: eigenvector cond llm}) becomes
\begin{equation}
    \frac{1-\lambda_y^2}{\phi_1} \big( \lambda_y^{2m_1}(\phi_2+\phi_3) + \lambda_y^{2m_2}\phi_3 \big) =
    \frac{1-\lambda_y^2}{\phi_2} \big( \lambda_y^{2m_1}(\phi_3+\phi_1) + \lambda_y^{2m_2}\phi_2 \big) =
    \frac{1-\lambda_y^2}{\phi_3} \big( \lambda_y^{2m_1}(\phi_1+\phi_2) + \lambda_y^{2m_2}\phi_1 \big).
\end{equation}
Multiplying by $\phi_1 \phi_2 \phi_3$, we have a polynomial equation of degree three, which therefore has exactly three solutions. Solving this equation for each case gives three solutions per case, and so $12$ eigenvectors in total, as required.

In general, we will have $N$ forms the eigenvectors can take, each with $|\Tilde{\mathcal{P}}|$ parameters to determine. Note that satisfying this equation is equivalent to solving a polynomial equation in $|\Tilde{\mathcal{P}}|$ variables. In this way, we can determine all of the eigenvectors of $\rho_y$, which we can then use to calculate $\Gamma(\mathcal{M})$, using Eq.~(\ref{eq: gamma expr}), and hence use Eqs.~(\ref{eq: a neq b gen expr N modes}) and (\ref{eq: a=b gen expr N modes}).

One special case we note is that for multisets $\mathcal{M}$ containing only a single unique element, e.g. $\mathcal{M}=\{m_1,m_1\}$, we do not need to solve any equation to determine $\Bar{\eta}$, as all of the eigenvectors are given directly by Eq.~(\ref{eq: phase relation}). E.g. for the $\mathcal{M}=\{m_1,m_1\}$ case, the valid eigenvectors are represented by $\bar{\eta}^{k=0} = \frac{1}{\sqrt{3}}(1,1,1)$, $\bar{\eta}^{k=1} = \frac{1}{\sqrt{3}}(1,e^{i\frac{2\pi}{3}},e^{i\frac{4\pi}{3}})$, and $\bar{\eta}^{k=2} = \frac{1}{\sqrt{3}}(1,e^{i\frac{4\pi}{3}},e^{i\frac{2\pi}{3}})$.

Finally, we note that whilst the method we present here allows the eigendecomposition of $\rho_y$ to be found analytically, for general $\lambda_y$ and $\mathcal{M}$, in practice this is likely to be impossible for large $N$ (or even for $N>4$). However, it is possible that there are significant simplifications possible that would allow the eigendecomposition for large $N$ to be found much more easily. In particular, the correspondence between the different categories of $\mathcal{M}$ that we need to consider and the number of ways of partitioning $N-1$ suggests a possible link to representation theory, which could be of use here.

\subsection{Numerical calculations of the channel output}\label{app: N port numerics}

In order to analytically calculate the channel output for an arbitrary coherent state, we need to find an analytical form for the eigenvectors of $\rho_y$ in terms of $\lambda_y$ and $\mathcal{M}$. In general, this is difficult. As shown in Subsection~\ref{app: N port eigendecomposition}, this can be done by solving a polynomial condition. However, this rapidly becomes impossible for large $N$. Since there are infinite values of $m_1$, $m_2$, etc, we cannot numerically find the eigenvectors for every multiset $\mathcal{M}$.

Nonetheless, Eqs.~(\ref{eq: a neq b gen expr N modes}) and (\ref{eq: a=b gen expr N modes}) make it very simple to numerically calculate the channel output with a truncation at some maximum value of $\mathcal{M}$. That is, instead of summing over every possible multiset of size $N-1$ integers, we sum over only those multisets whose largest elements are at most $M$. We now explain step by step how one can numerically calculate the channel output for an input element $\left|a\middle>\middle<b\right|$. If we have an arbitrary input state, we can express it as $\sum_{a,b} c_{a,b}\left|a\middle>\middle<b\right|$ for some set of coefficients $c_{a,b}$, so for a suitable truncation of the input state, we can also calculate the output for an arbitrary input.

From Eq.~(\ref{eq: a neq b gen expr N modes}), we know that we only need to calculate the magnitude of the term $\left|a\middle>\middle<b\right|$ in the output. The prefactor (outside of the brackets) is also simple to calculate, so we will focus on the sum within the brackets.

\begin{enumerate}
    \item Pick a distinct multiset $\mathcal{M}$ consisting of $N-1$ positive (or zero) integers, all of which are less than or equal to the maximum value $M$.
    \item Calculate the eigenvectors of $\rho_y$ corresponding to the multiset $\mathcal{M}$, $\{\left|\psi(\mathcal{M},i)\right>\}$, as per Section~\ref{app: N port eigendecomposition}. Equivalently, we find the set of orthonormal vectors $\{\eta_{\mathcal{M}}^{(i)}\}$ whose elements give the coefficients of the terms in the eigenvectors, per Eq.~(\ref{eq: eigenstate gen}). Also find the corresponding eigenvalues.
    \item Calculate the matrix $\Gamma(\mathcal{M})$ using the eigenvalues and eigenvectors, per Eq.~(\ref{eq: gamma expr}).
    \item Sum the elements of $\mathcal{M}$, and therefore calculate $\lambda_x^{2\sum_{m\in \mathcal{M}} m}$.
    \item Per Eq.~(\ref{eq: a neq b gen expr N modes}), check if either $a$ or $b$ is an element of $\mathcal{M}$. Depending on whether one or both is an element of $\mathcal{M}$, we carry out one of the four possible sums of specific elements of $\Gamma(\mathcal{M})$ and multiply the result by $\lambda_x^{2\sum_{m\in \mathcal{M}} m}$ to obtain the contribution to the sum from the multiset $\mathcal{M}$.
    \item Repeat from item 1 until we have summed over all multisets whose elements are all less than or equal to $M$.
\end{enumerate}

To illustrate, consider the case of the multiset $\mathcal{M}=\{m_1,m_1,m_2\}$ from the previous subsection, and suppose all of our eigenvectors were written in the form of Eq.~(\ref{eq: example eigenvector}). Then, if neither $a$ nor $b$ were equal to either $m_1$ or $m_2$, we would sum the on-diagonal elements of $\Gamma(\mathcal{M})$ in rows $1$, $5$, and $9$, i.e. those corresponding to terms in Eq.~(\ref{eq: example eigenvector}) where $n$ is in the first position. The contribution to the sum from this multiset would be $\lambda_x^{2(2m_1+m_2)}\sum_{i=\{1,5,9\}}\Gamma(\{m_1,m_1,m_2\})_{i,i}$. If instead we had the multiset $\mathcal{M}=\{m_1,m_1,b\}$, we would sum the elements $\Gamma(\{m_1,m_1,a\})_{1,1}$, $\Gamma(\{m_1,m_1,a\})_{12,1}$, $\Gamma(\{m_1,m_1,a\})_{5,5}$, $\Gamma(\{m_1,m_1,a\})_{7,5}$, $\Gamma(\{m_1,m_1,a\})_{9,9}$, and $\Gamma(\{m_1,m_1,a\})_{2,9}$ instead, i.e. the same rows, but with the columns corresponding to both the terms in Eq.~(\ref{eq: example eigenvector}) where $n$ is in the first position and those terms in which $n$ and $m_2$ have swapped positions.

We note that step 2 is likely to be the most computationally expensive step, for any $N>3$. If we had analytical expressions for the vectors $\{\eta_{\mathcal{M}}^{(i)}\}$ into which we could simply plug in the values of the elements of the multiset $\mathcal{M}$, this step would be easy. However, the method detailed in Section~\ref{app: N port eigendecomposition} requires solving polynomial equations of - in the worst case, in which every element of $\mathcal{M}$ is unique - degree $(N-1)!$ (and with $(N-1)!$ parameters). For polynomials of degree more than four, there is no explicit formula for the solution, and so even for $N=4$, this step involves numerically solving these polynomial equations for each $\mathcal{M}$ individually.

In the next Appendix, we will show that we can obtain analytical expressions for the vectors $\{\eta_{\mathcal{M}}^{(i)}\}$ for the $N=3$ case, and so all of the steps become mathematically simple. Nonetheless, Eqs.~(\ref{eq: a neq b gen expr N modes}) and (\ref{eq: a=b gen expr N modes}) (and the resulting three port expressions, Eqs.~(\ref{eq: n port output breakdown}), (\ref{eq: mm a neq b output}), (\ref{eq: mm a=b output}), (\ref{eq: lm a neq b output}), and (\ref{eq: lm a=b output})) still involve an infinite sum over all two-element multisets. Thus, in our example in Appendix~(\ref{app: TMSV}), we will use numerical calculations. The code used to numerically calculate the channel output for an arbitrary input state is available as supplementary MATLAB programs.

\section{Three port case}\label{app: 3 ports}

Let us explicitly apply the $N$ port formalism to the $N=3$ case. Our first task is to find the eigenvalues/eigenvectors corresponding to every two element multiset $\mathcal{M}$. We can split all of the possible multisets into two categories: multisets of the form $\{m,m\}$ (i.e. where both elements are the same) and multisets of the form $\{l,m\}$ (i.e. where the two elements are different). We will refer to the eigenvalues/eigenvectors corresponding to the first type of multiset as $\{mm\}$-eigenvalues/eigenvectors and those corresponding to the second type of multiset as $\{lm\}$-eigenvalues/eigenvectors.

$\{mm\}$-eigenvectors take the form
\begin{equation}
    \left|\psi(\{mm\},j)\right>_{CA} = \sqrt{\frac{1-\lambda_y^2}{\Xi_{\{mm\}}^{(j)}}}\sum_{n=0}^{\infty} (-\lambda_y)^n \left|n\right>_C \otimes \big( \eta^{(j)}_{1}\left|n m m\right> + \eta^{(j)}_{2}\left|m n m\right> + \eta^{(j)}_{3}\left|m m n\right> \big)_A,
    \label{eq: mm-eigenvector}
\end{equation}
and are labelled by vectors $\eta^{(j)}=(\eta^{(j)}_{1},\eta^{(j)}_{2},\eta^{(j)}_{3})$. Each multiset $\{mm\}$ (where $m$ is any non-negative integer) corresponds to three eigenvectors. The labelling vectors for these eigenvectors are
\begin{equation}
    \eta^{(1)}=\frac{1}{\sqrt{3}}(1,1,1),\quad
    \eta^{(2)}=\frac{1}{\sqrt{3}}(1,e^{i\frac{2\pi}{3}},e^{i\frac{4\pi}{3}}),\quad
    \eta^{(3)}=\frac{1}{\sqrt{3}}(1,e^{i\frac{4\pi}{3}},e^{i\frac{2\pi}{3}}),
\end{equation}
where the indexing of the different vectors $\eta^{(j)}$ is again arbitrary. The corresponding eigenvalues are
\begin{equation}
    \Xi_{\{mm\}}^{(1)} = 1+2(1-\lambda_y^2)\lambda_y^{2m},\quad
    \Xi_{\{mm\}}^{(2)} = \Xi_{\{mm\}}^{(3)} = 1-(1-\lambda_y^2)\lambda_y^{2m}.
\end{equation}

$\{lm\}$-eigenvectors take the form
\begin{equation}
    \begin{split}
        \left|\psi(\{lm\},j)\right>_{CA} = \sqrt{\frac{1-\lambda_y^2}{\Xi_{\{lm\}}^{(j)}}}\sum_{n=0}^{\infty} (-\lambda_y)^n \left|n\right>_C \otimes \big(& \eta^{(j)}_{1}\left|n l m\right> + \eta^{(j)}_{2}\left|l m n\right> + \eta^{(j)}_{3}\left|m n l\right>\\
        &+ \eta^{(j)}_{4}\left|n m l\right> + \eta^{(j)}_{5}\left|m l n\right> + \eta^{(j)}_{6}\left|l n m\right> \big)_A,
    \end{split}
    \label{eq: lm-eigenvector}
\end{equation}
and the six labelling vectors, $\eta^{(j)}=(\eta^{(j)}_{1},\eta^{(j)}_{2},\eta^{(j)}_{3},\eta^{(j)}_{4},\eta^{(j)}_{5},\eta^{(j)}_{6})$, are
\begin{equation}
    \begin{split}
        &\eta^{(1)}=\frac{1}{\sqrt{6}}(1,1,1,1,1,1),\quad
        \eta^{(2)}=\frac{1}{\sqrt{6}}(1,1,1,-1,-1,-1),\\
        &\eta^{(3)}=\frac{1}{\sqrt{6}}(1,e^{i\frac{2\pi}{3}},e^{i\frac{4\pi}{3}},e^{i\phi},e^{i(\frac{2\pi}{3}+\phi)},e^{i(\frac{4\pi}{3}+\phi)},\quad
        \eta^{(4)}=\frac{1}{\sqrt{6}}(1,e^{i\frac{2\pi}{3}},e^{i\frac{4\pi}{3}},-e^{i\phi},-e^{i(\frac{2\pi}{3}+\phi)},-e^{i(\frac{4\pi}{3}+\phi)},\\
        &\eta^{(5)}=\frac{1}{\sqrt{6}}(1,e^{i\frac{4\pi}{3}},e^{i\frac{2\pi}{3}},e^{-i\phi},e^{i(\frac{4\pi}{3}-\phi)},e^{i(\frac{2\pi}{3}-\phi)},\quad
        \eta^{(6)}=\frac{1}{\sqrt{6}}(1,e^{i\frac{4\pi}{3}},e^{i\frac{2\pi}{3}},-e^{-i\phi},-e^{i(\frac{4\pi}{3}-\phi)},-e^{i(\frac{2\pi}{3}-\phi)},
    \end{split}\label{eq: lm vectors}
\end{equation}
where $\phi$ takes the real value $\phi = \frac{4\pi}{3} - \arg[1+\lambda_y^{2(l-m)}e^{i\frac{2\pi}{3}}]$. The resulting eigenvalues are
\begin{equation}
    \begin{split}
        &\Xi_{\{lm\}}^{(1)} = 1+(1-\lambda_y^2)(\lambda_y^{2l}+\lambda_y^{2m}),\quad
        \Xi_{\{lm\}}^{(2)} = 1-(1-\lambda_y^2)(\lambda_y^{2l}+\lambda_y^{2m}),\\
        &\Xi_{\{lm\}}^{(3)} = \Xi_{\{lm\}}^{(5)} = 1+(1-\lambda_y^2)\sqrt{\lambda_y^{4l}-\lambda_y^{2(l+m)}+\lambda_y^{4m}},\\
        &\Xi_{\{lm\}}^{(4)} = \Xi_{\{lm\}}^{(6)} = 1-(1-\lambda_y^2)\sqrt{\lambda_y^{4l}-\lambda_y^{2(l+m)}+\lambda_y^{4m}}.
    \end{split}\label{eq: lm eigenvalues}
\end{equation}

The operator $\rho_y$ can therefore be expressed as
\begin{equation}
    \rho_y = \sum_{m=0}^{\infty} \sum_{j=1}^{3}  \Xi_{\{mm\}}^{(j)} \left|\psi(\{mm\},j)\middle>\middle<\psi(\{mm\},j)\right| + \sum_{\substack{l,m=0,\\l > m}}^{\infty} \sum_{j=1}^{6}  \Xi_{\{lm\}}^{(j)} \left|\psi(\{lm\},j)\middle>\middle<\psi(\{lm\},j)\right|,
\end{equation}
where we have separated out the contributions from the $\{mm\}$-eigenvectors and the $\{lm\}$-eigenvectors. We note that since some of the eigenvalues are degenerate, this decomposition is not unique.

Next, we must calculate $\Gamma(\mathcal{M})$ for each multiset, according to Eq.~(\ref{eq: gamma expr}). The only part of this equation that may not be immediately clear is the meaning of the term $\sum_{\Phi\in\Tilde{\mathcal{P}}}\eta_{\mathcal{M},\Phi}^{(\alpha)*}\eta_{\mathcal{M},\Phi}^{(\beta)}$. However, now that we have fixed a labelling convention for the vectors $\{\eta^{(j)}\}$, we can more simply explain which elements of the vectors each $\Phi$ is labelling. $\Tilde{\mathcal{P}}$ is the set of indices corresponding to the coefficients of the component states in each eigenvector for which $n$ is in system $A_1$. For the $\{mm\}$-eigenvectors, this means it is only the index $1$, since $\eta^{(j)}_{1}$ is the coefficient for $\left|nmm\right>$ in Eq.~(\ref{eq: mm-eigenvector}). For the $\{lm\}$-eigenvectors, $\Tilde{\mathcal{P}}$ contains the indices $1$ and $4$ (corresponding to $\left|nlm\right>$ and $\left|nml\right>$ respectively, per Eq.~(\ref{eq: lm-eigenvector})).

For the $\{mm\}$-eigenvectors, $\sum_{\Phi\in\Tilde{\mathcal{P}}}\eta_{\mathcal{M},\Phi}^{(\alpha)*}\eta_{\mathcal{M},\Phi}^{(\beta)}=\frac{1}{3}$ for every $\alpha$ and $\beta$, whilst for the $\{lm\}$-eigenvectors, it gives $\frac{1}{6}(1+e^{\pm i\phi})$, $\frac{1}{6}(1-e^{\pm i\phi})$, $\frac{1}{6}(1+e^{\pm 2i\phi})$, or $0$, depending on the values of $\alpha$ and $\beta$. For the $\{mm\}$ case, we calculate
\begin{equation}
    \Gamma(\{m,m\})=\frac{1}{9}\left( \frac{1}{\sqrt{\Xi_{\{mm\}}^{(1)}\Xi_{\{mm\}}^{(2)}}}\begin{pmatrix}
        4 &1 &1\\
        1 &-2 &-2\\
        1 &-2 &-2
    \end{pmatrix}
    + \frac{1}{\Xi_{\{mm\}}^{(2)}}\begin{pmatrix}
        2 &-1 &-1\\
        -1 &-1 &2\\
        -1 &2 &-1
    \end{pmatrix}\right).\label{eq: gamma mm}
\end{equation}
In the $\{lm\}$ case, we calculate (the full expression is unwieldy, but can be found in the supplementary Mathematica notebook)
\begin{equation}
    \begin{split}
        \Gamma(\{l,m\}) = \gamma + \gamma^{\dagger},\quad 
        \gamma = \frac{1}{6}\Bigg(& \frac{(1+e^{i\phi})(\eta^{(1)\dagger}\eta^{(3)}+\eta^{(5)\dagger}\eta^{(1)})}{\sqrt{\Xi_{\{lm\}}^{(1)}\Xi_{\{lm\}}^{(3)}}} +
        \frac{(1-e^{i\phi})(\eta^{(1)\dagger}\eta^{(4)}+\eta^{(6)\dagger}\eta^{(1)})}{\sqrt{\Xi_{\{lm\}}^{(1)}\Xi_{\{lm\}}^{(4)}}} +\\
        &\frac{(1-e^{i\phi})(\eta^{(2)\dagger}\eta^{(3)}+\eta^{(5)\dagger}\eta^{(2)})}{\sqrt{\Xi_{\{lm\}}^{(2)}\Xi_{\{lm\}}^{(3)}}} +
        \frac{(1+e^{i\phi})(\eta^{(2)\dagger}\eta^{(4)}+\eta^{(6)\dagger}\eta^{(2)})}{\sqrt{\Xi_{\{lm\}}^{(2)}\Xi_{\{lm\}}^{(4)}}} +\\
        &\frac{(1+e^{2i\phi})\eta^{(5)\dagger}\eta^{(3)}}{\Xi_{\{lm\}}^{(3)}} +
        \frac{(1-e^{2i\phi})\eta^{(6)\dagger}\eta^{(3)}}{\sqrt{\Xi_{\{lm\}}^{(3)}\Xi_{\{lm\}}^{(4)}}} +
        \frac{(1+e^{2i\phi})\eta^{(4)\dagger}\eta^{(6)}}{\Xi_{\{lm\}}^{(4)}}
        \Bigg).
    \end{split}\label{eq: gamma lm}
\end{equation}

We may now begin applying Eqs.~(\ref{eq: a neq b gen expr N modes}) and (\ref{eq: a=b gen expr N modes}). It is helpful to split each equation into the contributions from multisets of the form $\{mm\}$ and from multisets of the form $\{lm\}$, so that
\begin{equation}
    \mathcal{E}_{x,y}[\left|a\middle>\middle<b\right|] = \delta_{ab}(1-\lambda_x^2)\sum_{n=0}^{\infty} \lambda_x^{2n} \left|n\middle>\middle<n\right|
    + \mathcal{E}^{\{mm\}}_{x,y}[\left|a\middle>\middle<b\right|] + \mathcal{E}^{\{lm\}}_{x,y}[\left|a\middle>\middle<b\right|].\label{eq: n port output breakdown}
\end{equation}
Then, applying Eq.~(\ref{eq: a neq b gen expr N modes}), we find
\begin{equation}
    \begin{split}
        \mathcal{E}^{\{mm\}}_{x,y}[\left|a\middle>\middle<b\right|] = 3 (1-\lambda_x^2)^{3}&(1-\lambda_y^2)(\lambda_x \lambda_y)^{a+b}
        \bigg(\sum_{m=0}^{\infty} \lambda_x^{4m}
        \Gamma(\{m,m\})_{1,1} +\\
        &\lambda_x^{4a} \sum_{i=\{2,3\}} \Gamma(\{a,a\})_{1,i} +
        \lambda_x^{4b} \sum_{i=\{2,3\}} \Gamma(\{b,b\})_{i,1}\bigg) \left|a\middle>\middle<b\right|,
    \end{split}
\end{equation}
and applying Eq.~(\ref{eq: a=b gen expr N modes}), we get
\begin{equation}
    \begin{split}
        \mathcal{E}^{\{mm\}}_{x,y}[\left|a\middle>\middle<a\right|] = 3(1-\lambda_x^2)^{3}(1-\lambda_y^2) (\lambda_x\lambda_y)^{2a} \Bigg(\bigg( \sum_{m=0,m\neq a}
        \lambda_x^{4m} \Gamma(\{m,m\})_{1,1} +
        \lambda_x^{4a} \sum_{i,j=1}^3 \Gamma(\{a,a\})_{i,j} \bigg) \left|a\middle>\middle<a\right| +\\
        \sum_{m=0,m\neq a}
        \lambda_x^{4m} \sum_{i=\{2,3\}} \Gamma(\{m,m\})_{i,i} \left|m\middle>\middle<m\right| \Bigg).
    \end{split}
\end{equation}
However, since we have the explicit expression for $\Gamma(\{m,m\})$, we can simplify these expressions further, getting
\begin{align}
    &\mathcal{E}^{\{mm\}}_{x,y}[\left|a\middle>\middle<b\right|] = 3 (1-\lambda_x^2)^{3}(1-\lambda_y^2)(\lambda_x \lambda_y)^{a+b}
    \bigg(\sum_{m=0}^{\infty} \lambda_x^{4m}
    \Gamma(\{m,m\})_{1,1} +
    2\lambda_x^{4a} \Gamma(\{a,a\})_{1,2} + 2\lambda_x^{4b} \Gamma(\{b,b\})_{1,2}\bigg) \left|a\middle>\middle<b\right|,\label{eq: mm a neq b output}\\
    &\mathcal{E}^{\{mm\}}_{x,y}[\left|a\middle>\middle<a\right|] = 3(1-\lambda_x^2)^{3}(1-\lambda_y^2) (\lambda_x\lambda_y)^{2a} \sum_{m=0,m\neq a}
    \lambda_x^{4m} \Gamma(\{m,m\})_{1,1} (\left|a\middle>\middle<a\right| - \left|m\middle>\middle<m\right|),\label{eq: mm a=b output}
\end{align}
so we only need two elements of $\Gamma(\{m,m\})$. In the $\{lm\}$ case, Eq.~(\ref{eq: a neq b gen expr N modes}) gives us
\begin{equation}
    \begin{split}
        \mathcal{E}^{\{lm\}}_{x,y}[\left|a\middle>\middle<b\right|] = \frac{3}{2} &(1-\lambda_x^2)^{3}(1-\lambda_y^2)(\lambda_x \lambda_y)^{a+b}
        \Bigg(\sum_{\substack{l,m=0,\\l\neq m}}^{\infty} \lambda_x^{2(l+m)}
        \sum_{i=\{1,4\}} \Gamma(\{l,m\})_{i,i} +\\
        &\sum_{\substack{m=0,\\m\neq a}} \lambda_x^{2(a+m)}
        \Big( \Gamma(\{a,m\})_{1,6} + \Gamma(\{a,m\})_{4,2} \Big) +
        \sum_{\substack{l=0,\\l\neq a}} \lambda_x^{2(a+l)}
        \Big( \Gamma(\{l,a\})_{1,5} + \Gamma(\{l,a\})_{4,3} \Big) +\\
        &\sum_{\substack{m=0,\\m\neq b}} \lambda_x^{2(b+m)}
        \Big( \Gamma(\{b,m\})_{6,1} + \Gamma(\{b,m\})_{2,4} \Big) +
        \sum_{\substack{l=0,\\l\neq b}} \lambda_x^{2(b+l)}
        \Big( \Gamma(\{l,b\})_{5,1} + \Gamma(\{l,b\})_{3,4} \Big) +\\
        &\lambda_x^{2(a+b)} \big( \Gamma(\{a,b\})_{5,6} + \Gamma(\{a,b\})_{3,2} + \Gamma(\{b,a\})_{6,5} + \Gamma(\{b,a\})_{2,3} \big)
        \Bigg) \left|a\middle>\middle<b\right|,
    \end{split}
\end{equation}
where the factor of $\frac{1}{2}$ is because we double count each term $\{l,m\}$ by also summing over $\{m,l\}$. We now note that if we send $\{l,m\}$ to $\{m,l\}$, then we send $\phi$ to $-\phi$. This does not change the eigenvalues in Eq.~(\ref{eq: lm eigenvalues}) but changes the vectors in Eq.~(\ref{eq: lm vectors}) in the following way: $\eta^{(3)}$ becomes $\eta^{(5)}$, $\eta^{(4)}$ becomes $\eta^{(6)}$, and vice versa, and the second and third elements of each vector are swapped, along with the fifth and sixth elements. Propagating this change to the expression in Eq.~(\ref{eq: gamma lm}), we see that $\Gamma(\{l,m\})_{1,5}=\Gamma(\{m,l\})_{6,1}$, $\Gamma(\{l,m\})_{4,3}=\Gamma(\{m,l\})_{2,4}$, $\Gamma(\{l,m\})_{5,6}=\Gamma(\{m,l\})_{5,6}$, and $\Gamma(\{l,m\})_{3,2}=\Gamma(\{m,l\})_{3,2}$. Using also the fact that $\Gamma(\mathcal{M})$ is hermitian by construction, we get
\begin{equation}
    \begin{split}
        \mathcal{E}^{\{lm\}}_{x,y}[\left|a\middle>\middle<b\right|] =& 3 (1-\lambda_x^2)^{3}(1-\lambda_y^2)(\lambda_x \lambda_y)^{a+b}
        \Bigg(\sum_{m=0}^{\infty} \lambda_x^{2m}\bigg( \frac{1}{2} \sum_{l=0,l\neq m}^{\infty} \lambda_x^{2l}
        \Big( \Gamma(\{l,m\})_{1,1} + \Gamma(\{l,m\})_{4,4} \Big) +\\
        &(1-\delta_{am}) \lambda_x^{2a}
        \mathrm{Re}\big[ \Gamma(\{a,m\})_{1,6} + \Gamma(\{a,m\})_{4,2} \big] +\\
        &(1-\delta_{bm}) \lambda_x^{2b}
        \mathrm{Re}\big[ \Gamma(\{b,m\})_{1,6} + \Gamma(\{b,m\})_{4,2} \big] \bigg) +
        \lambda_x^{2(a+b)} \mathrm{Re}\big[ \Gamma(\{a,b\})_{5,6} + \Gamma(\{a,b\})_{3,2} \big]
        \Bigg) \left|a\middle>\middle<b\right|.
    \end{split}\label{eq: lm a neq b output}
\end{equation}
Note that, from Eqs.~(\ref{eq: mm a neq b output}) and (\ref{eq: lm a neq b output}), the scaling factor for the number state element $\left|a\middle>\middle<b\right|$ is always real. Eq.~(\ref{eq: a=b gen expr N modes}) gives us
\begin{equation}
    \begin{split}
        \mathcal{E}^{\{lm\}}_{x,y}[\left|a\middle>\middle<a\right|] =&
        3 (1-\lambda_x^2)^{3}(1-\lambda_y^2) (\lambda_x\lambda_y)^{2a} \Bigg( \sum_{m=0}^{\infty} \lambda_x^{2m}\bigg( \frac{1}{2} \sum_{\substack{l=0,\\ l\neq m}}^{\infty}
        \lambda_x^{2l} \Big( \Gamma(\{l,m\})_{1,1} + \Gamma(\{l,m\})_{4,4} \Big) +\\
        &(1-\delta_{am}) \lambda_x^{2a} \Big( 2\mathrm{Re}[\Gamma(\{a,m\})_{1,6} + \Gamma(\{a,m\})_{4,2}] + \Gamma(\{a,m\})_{2,2} + \Gamma(\{a,m\})_{6,6} \Big)
        \bigg) \left|a\middle>\middle<a\right| +\\
        &\sum_{\substack{m=0,\\ m\neq a}}^{\infty} \lambda_x^{2m} \bigg( \sum_{\substack{l=0,\\ l\neq m}}^{\infty}
        \lambda_x^{2l} \Big( \Gamma(\{l,m\})_{3,3} + \Gamma(\{l,m\})_{5,5} \Big) +
        2 \lambda_x^{2a} \mathrm{Re}[\Gamma(\{a,m\})_{3,5}] \bigg)\left|m\middle>\middle<m\right| \Bigg),
    \end{split}
\end{equation}
which reduces to
\begin{equation}
    \begin{split}
        \mathcal{E}^{\{lm\}}_{x,y}[\left|a\middle>\middle<a\right|] =&
        3 (1-\lambda_x^2)^{3}(1-\lambda_y^2) (\lambda_x\lambda_y)^{2a} \Bigg( \sum_{m=0}^{\infty} \lambda_x^{2m}\bigg( \frac{1}{2} \sum_{\substack{l=0,\\ l\neq m}}^{\infty}
        \lambda_x^{2l} \Big( \Gamma(\{l,m\})_{1,1} + \Gamma(\{l,m\})_{4,4} \Big) +\\
        &(1-\delta_{am}) 2\lambda_x^{2a} \mathrm{Re}[\Gamma(\{a,m\})_{1,6} + \Gamma(\{a,m\})_{4,2}]
        \bigg) \left|a\middle>\middle<a\right| +\\
        &\sum_{m=0}^{\infty} \lambda_x^{2m} \bigg( \sum_{\substack{l=0,\\ l\neq m}}^{\infty}
        \lambda_x^{2l} \Big( \Gamma(\{l,m\})_{3,3} + \Gamma(\{l,m\})_{5,5} \Big) +
        (1-\delta_{am}) 2 \lambda_x^{2a} \mathrm{Re}[\Gamma(\{a,m\})_{3,5}] \bigg)\left|m\middle>\middle<m\right| \Bigg).
    \end{split}\label{eq: lm a=b output}
\end{equation}

Thus, by combining Eqs.~(\ref{eq: n port output breakdown}), (\ref{eq: mm a neq b output}), (\ref{eq: mm a=b output}), (\ref{eq: lm a neq b output}), and (\ref{eq: lm a=b output}), we have an expression for the channel enacted on any number state element $\left|a\middle>\middle<b\right|$ when we carry out three port CV-PBT using TMSVs with squeezing parameter $\lambda_x$ as a resource. These expressions can be easily numerically evaluated, as we will show in the next Appendix.

\section{Applying CV-PBT to a TMSV and Bell states}\label{app: TMSV}

We will now apply the expressions in Eqs.~(\ref{eq: n port output breakdown}), (\ref{eq: mm a neq b output}), (\ref{eq: mm a=b output}), (\ref{eq: lm a neq b output}), and (\ref{eq: lm a=b output}) to calculate the output state for three different types of entangled input state, as a demonstration of how to numerically calculate the channel output.

The first scenario we look at is sending one mode of a TMSV using three port CV-PBT. Letting $\lambda_{\mathrm{in}}$ parameterise the degree of entanglement of the input state, we recall that the state we want to send takes the form
\begin{equation}
    \psi_{\mathrm{in}}(\lambda_{\mathrm{in}}) = (1-\lambda_{\mathrm{in}}^2)\sum_{a,b=0}^{\infty} (-\lambda_{\mathrm{in}})^{a+b} \left|aa\middle>\middle<bb\right|_{SI},\label{eq: input TMSV}
\end{equation}
where $S$ is the signal state that we wish to teleport and $I$ is the idler system. We wish to calculate
\begin{equation}
    \psi_{\mathrm{out}}(\lambda_{\mathrm{in}}) = (1-\lambda_{\mathrm{in}}^2)\sum_{a,b=0}^{\infty} (-\lambda_{\mathrm{in}})^{a+b} \mathcal{E}_{x,y}[\left|a\middle>\middle<b\right|]_S \otimes \left|a\middle>\middle<b\right|_{I},
\end{equation}
and per Eq.~(\ref{eq: n port output breakdown}), this is given by
\begin{equation}
    \psi_{\mathrm{out}}(\lambda_{\mathrm{in}}) = (1-\lambda_{\mathrm{in}}^2)\sum_{a,b=0}^{\infty} (-\lambda_{\mathrm{in}})^{a+b} \left( \delta_{ab}(1-\lambda_x^2)\sum_{n=0}^{\infty} \lambda_x^{2n} \left|n\middle>\middle<n\right|
    + \mathcal{E}^{\{mm\}}_{x,y}[\left|a\middle>\middle<b\right|] + \mathcal{E}^{\{lm\}}_{x,y}[\left|a\middle>\middle<b\right|] \right)_S \otimes \left|a\middle>\middle<b\right|_{I}.\label{eq: TMSV output}
\end{equation}

There are two different types of truncation we must apply. First, we must apply a truncation to the output state by choosing a maximum value of $a$ and $b$ that we will calculate up to, since we are dealing with infinite-dimensional systems. Secondly, when calculating the term $\mathcal{E}^{\{mm\}}_{x,y}[\left|a\middle>\middle<b\right|]$ (using Eqs.~(\ref{eq: mm a neq b output}) and (\ref{eq: mm a=b output})), we must choose a maximum value of $m$ to use in the sums, and when calculating the term $\mathcal{E}^{\{lm\}}_{x,y}[\left|a\middle>\middle<b\right|]$ (using Eqs.~(\ref{eq: lm a neq b output}) and (\ref{eq: lm a=b output})), we must choose maximum values of $l$ and $m$. The first type of truncation affects how much of the output state we will have access to (i.e. how big a subset of the density matrix) but not the accuracy of the element that we do have, since (due to the idler system) terms corresponding to different $a$ and $b$ do not affect each other. The second type of truncation affects the accuracy, so the maximum values of $l$ and $m$ should be suitably large.

Numerical calculations of Eq.~(\ref{eq: TMSV output}) are implemented in the supplementary MATLAB code (as well as for the $N=2$ case). We also calculate the fidelity between the input and output states, as shown in Fig.~\ref{fig: tmsv} (in the main text). Note that we call the fidelity here, $F=\mathrm{Tr}\left[\sqrt{\sqrt{\sigma}\rho\sqrt{\sigma}}\right]^2$, is sometimes also called the squared fidelity.

Next, we consider applying the protocol to one half of a maximally entangled two qubit state (a Bell state). The input state is similar to the state in Eq.~(\ref{eq: input TMSV}), but we only have four terms and the prefactors for all of them are the same:
\begin{equation}
    \psi_{\mathrm{in}}^{(2)} = \frac{1}{2} \sum_{a,b=0}^{1} \left|aa\middle>\middle<bb\right|_{SI}.\label{eq: input Bell}
\end{equation}
This simplifies the calculation significantly, since we no longer need to choose maximum values for $a$ and $b$. Note that despite the fact our input state is DV, the output will still be CV (although the only components that lie outside of the four dimensional subspace will be on-diagonal components).

\begin{figure}[tb]
\vspace{+0.1cm}
\centering
\includegraphics[width=1\linewidth]{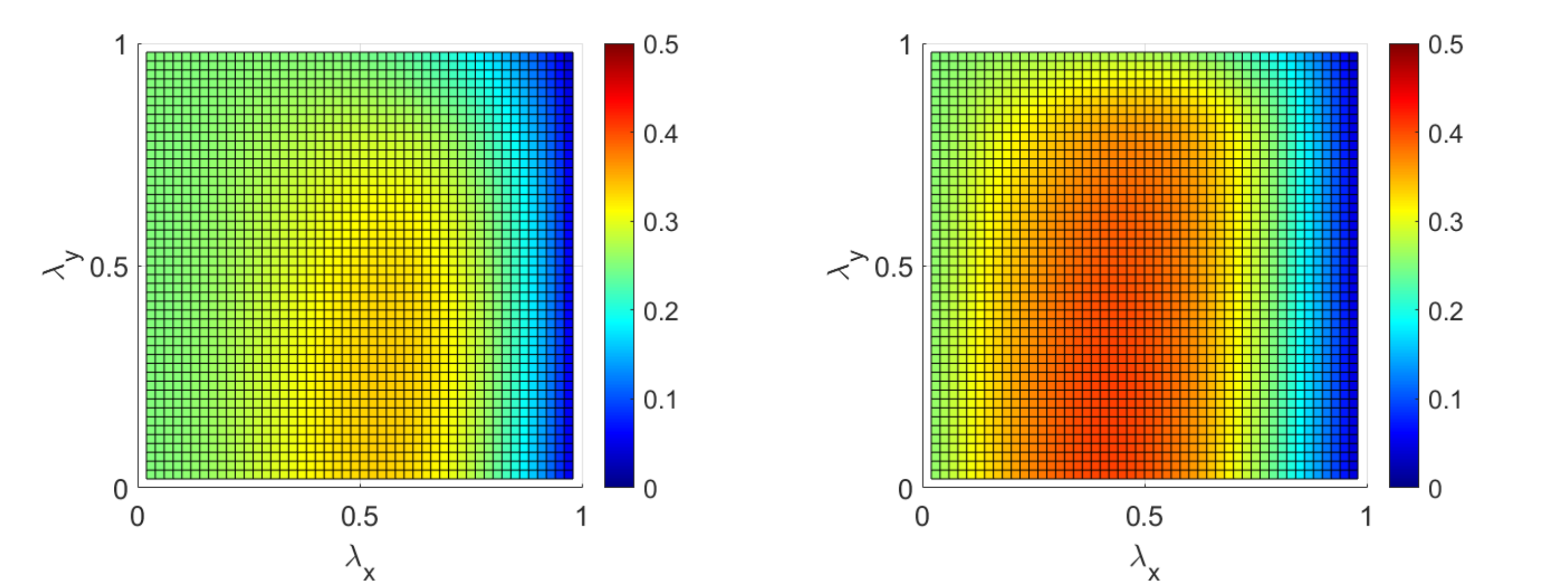}\caption{The input-output fidelity when one half of a maximally entangled qubit state is acted on with CV-PBT. The plot on the left shows the two port case, whilst the plot on the right shows the three port case.}
\label{fig: Bell2}
\end{figure}

The fidelity between the input and output states is plotted in Fig.~(\ref{fig: Bell2}) for both the $N=2$ and $N=3$ cases. Note that for this calculation, we can truncate our output state to a two qubit subspace without adding any error to our calculations, so that they are almost exact (although we still have maximum values of $l$ and $m$ in our sums).

Three port CV-PBT results in a better input-output fidelity than two port CV-PBT over a wide range of values, although neither has a very high fidelity. This is not surprising, since we are sending a very small DV system using a CV protocol. Also, the Bell state has the same prefactors for each component, whilst for the output from CV-PBT, the weightings of the on-diagonal states decay for larger energies. This is reflected in the fact that CV-PBT can send low energy TMSVs with a much higher input-output fidelity (as in Fig.~\ref{fig: tmsv}).

\begin{figure}[tb]
\vspace{+0.1cm}
\centering
\includegraphics[width=1\linewidth]{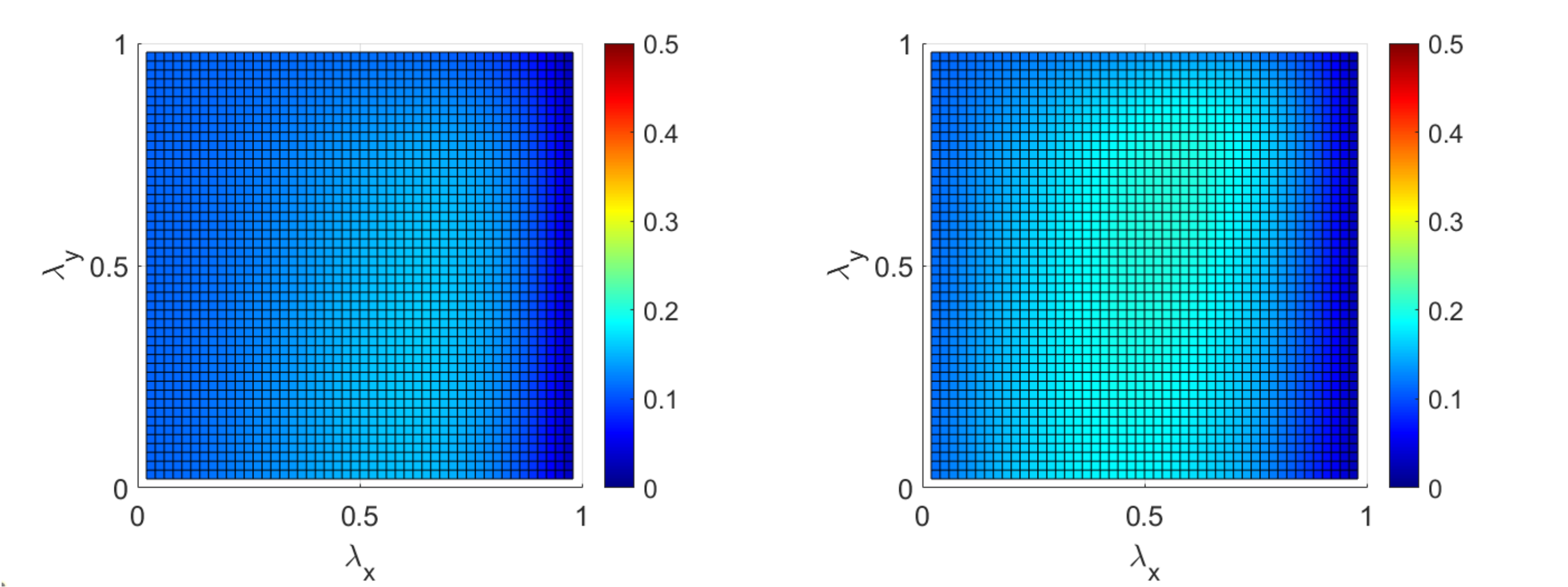}\caption{The input-output fidelity when one half of a maximally entangled qutrit state is acted on with CV-PBT. The plot on the left shows the two port case, whilst the plot on the right shows the three port case.}
\label{fig: Bell3}
\end{figure}

The final case is similar to the second case. We send a maximally entangled qutrit state through the channel. The fidelities are shown in Fig.~(\ref{fig: Bell3}). Again, three port CV-PBT results in a better input-output fidelity than two port CV-PBT over a wide range of values. For both, the fidelities are lower than for the maximally entangled qubit input. We can understand this in terms of the energy truncation. As shown in Fig.~\ref{fig: energy}, the maximum average photon number of the two port CV-PBT output is lower than one until $\lambda_x$ is large. For a maximally entangled qutrit state, the average energy of each system is $1$ (in the qubit case, it is $\frac{1}{2}$). Thus, two port CV-PBT is not even able to output a state with the same energy until $\lambda_x$ is large.

\end{document}